\documentclass[lettersize,journal]{IEEEtran}

\usepackage[utf8]{inputenc}
\usepackage[english]{babel}
\usepackage{hyperref}
\hypersetup{
    colorlinks=true,
    linkcolor=blue,
    anchorcolor=blue,
    filecolor=blue,      
    citecolor=blue,
    urlcolor=blue
}
\usepackage{longtable}
\usepackage{pdflscape}
\usepackage{lineno}
\usepackage{subcaption}
\usepackage{lipsum}
\usepackage{algorithm}
\usepackage{algpseudocode}
\usepackage{graphicx}
\usepackage[compatibility=false]{caption}
\usepackage{booktabs}
\usepackage{cite}
\usepackage{url}
\usepackage{multirow}
\usepackage{pgfplots}
\usepackage{tikz}
\usetikzlibrary{matrix,fit,shapes,calc,positioning,shadows,arrows,shapes,backgrounds,decorations.markings,fadings}
\usepgfplotslibrary{statistics}
\usepackage{listings}

\usepackage{flushend}
\usepackage{balance}
\usepackage{wrapfig}
\usepackage{enumitem}
\usepackage{csquotes}
\usepackage{color, colortbl}
\definecolor{Gray}{gray}{0.9}
\definecolor{cadmiumgreen}{rgb}{0.0, 0.42, 0.24}

\usepackage{multirow}
\usepackage[normalem]{ulem}
\useunder{\uline}{\ul}{}

\usepackage[skins]{tcolorbox}

\usepackage{xfp}
\usepackage{siunitx}

\newcommand{\etal}{et al.}  
\newcommand{\CodeIn}[1]{{\small{\texttt{#1}}}}

\newcommand{\MyComment}[1]{}

\newcommand{\OurComment}[1]{}
\newcommand{\Space}[1]{}
\newcommand{\Num}[1]{#1}
\newcommand{\MacroNum}[1]{#1}

\newcommand{\precaptionspace}{\vspace{0ex}}

\newcommand{\postcaptionsCodespace}{\vspace{-2ex}}

\newcommand{\Fix}[1]{{\textbf{[[}\color{magenta}#1}\textbf{]]}}

\newcommand{\Den}[1]{[\textbf{Denini}:~{\color{brown} #1}]}
\newcommand{\swnote}[1]{[\textbf{Stefan}:~{\color{cadmiumgreen} #1}]}
\newcommand{\jon}[1]{[\textbf{Jon}:~{\color{orange} #1}]}

\newcommand{\alexi}[1]{[\textbf{Alexi}:~{\color{blue} #1}]}

\newcommand{\cpuOnly}{{(C)}}
\newcommand{\memOnly}{{(M)}}
\newcommand{\diskOnly}{{(D)}}
\newcommand{\networkOnly}{{(N)}}
\newcommand{\cpuMem}{(CM)}
\newcommand{\cpuNetwork}{(CN)}
\newcommand{\memDisk}{(MD)}
\newcommand{\memNetwork}{(MN)}
\newcommand{\cpuDisk}{(CD)}
\newcommand{\diskNetwork}{(DN)}
\newcommand{\cpuMemNetwork}{(CMN)}
\newcommand{\cpuMemDisk}{(CMD)}
\newcommand{\cpuDiskNetwork}{(CDN)}
\newcommand{\memDiskNetwork}{(MDN)}
\newcommand{\cpuMemDiskNetwork}{(CMDN)}

\newcommand{\numProjJava}{\MacroNum{30}}
\newcommand{\numProjJS}{\MacroNum{10}}
\newcommand{\numProjPython}{\MacroNum{12}}
\newcommand{\numProj}{\MacroNum{52}}
\newcommand{\numProjWithoutRAFT}{\MacroNum{15}}
\newcommand{\numProjRAFTContact}{\MacroNum{14}}
\newcommand{\numProjRAFTPending}{\MacroNum{eight}}
\newcommand{\numProjRAFTResponded}{\MacroNum{six}}
\newcommand{\numProjRAFTRespondedFix}{\MacroNum{two}}
\newcommand{\numProjRAFTRespondedSpecs}{\MacroNum{four}}
\newcommand{\numConfigs}{\MacroNum{27}}
\newcommand{\numConfigsPhaseOne}{\MacroNum{16}}
\newcommand{\numFlakeFlaggerDataset}{\MacroNum{15}}
\newcommand{\numLamDataset}{\MacroNum{15}}
\newcommand{\numTestRuns}{\MacroNum{300}}
\newcommand{\numBaselineFlaky}{86}
\newcommand{\numTotalFlaky}{608}
\newcommand{\numTotalRAFT}{283}
\newcommand{\numTotalRAFTInOneConfig}{\MacroNum{24}}
\newcommand{\percentRAFT}{\fpeval{trunc((\numTotalRAFT / \numTotalFlaky) * 100, 1)}}
\newcommand{\numJavaWebsocketFlaky}{37}
\newcommand{\numJavaWebsocketRAFT}{32}
\newcommand{\percentJavaWebsocketRAFT}{\fpeval{trunc((\numJavaWebsocketRAFT / \numJavaWebsocketFlaky) * 100, 2)}}

\usepackage{booktabs,arydshln}
\makeatletter
\def\adl@drawiv#1#2#3{%
        \hskip.5\tabcolsep
        \xleaders#3{#2.5\@tempdimb #1{1}#2.5\@tempdimb}%
                #2\z@ plus1fil minus1fil\relax
        \hskip.5\tabcolsep}
\newcommand{\cdashlinelr}[1]{%
  \noalign{\vskip\aboverulesep
           \global\let\@dashdrawstore\adl@draw
           \global\let\adl@draw\adl@drawiv}
  \cdashline{#1}
  \noalign{\global\let\adl@draw\@dashdrawstore
           \vskip\belowrulesep}}

\newcommand{\rqone}{How prevalent are RAFTs?}
\newcommand{\rqoneone}{How many of the flaky test failures can be attributed to resource starvation?}
\newcommand{\rqonetwo}{How sensitive are RAFT failures to resource starvation?}

\newcommand{\rqtwo}{Which resources have the strongest influence on flakiness?}
\newcommand{\rqtwoone}{What resources are most common at triggering flaky test failures?}
 \newcommand{\rqtwotwo}{Are some flaky tests only detected when using different combinations of resources?}

\newcommand{\rqthree}{Which configuration best saves money while running the test suite \underline{\smash{to prevent RAFTs}}?}

\newcommand{\rqfour}{Which configuration best saves money while running the test suite \underline{\smash{to detect RAFTs}}?}

\definecolor{Green}{rgb}{0.6,1,0.8}
\definecolor{Red}{rgb}{0.99, 0.76, 0.8}

\definecolor{dkgreen}{rgb}{0,0.6,0}
\definecolor{gray}{rgb}{0.5,0.5,0.5}
\definecolor{mauve}{rgb}{0.58,0,0.82}

\begin{document}

\title{The Effects of Computational Resources on Flaky Tests}

\author{Denini Silva, Martin Gruber, Satyajit Gokhale, Ellen Arteca, Alexi Turcotte, Marcelo d'Amorim, Wing Lam, Stefan Winter, and Jonathan Bell%
\thanks{Denini Silva is with Federal University of Pernambuco, Recife, PE Brazil. (e-mail: dgs@cin.ufpe.br)}%
\thanks{Martin Gruber is with BMW Group and University of Passau, 94032 Passau, Germany. (e-mail: martin.gr.gruber@bmw.de)}%
\thanks{Satyajit Gokhale, Ellen Arteca, Alexi Torcotte and Jonathan Bell are with Northeastern University, Boston, MA 02215 USA. (e-mails: \{gokhale.sa, arteca.e, turcotte.al, jbell\}@northeastern.edu)}%
\thanks{Marcelo d'Amorim is with North Carolina State University, Raleigh, NC 27695 USA. (e-mail: mdamori@ncsu.edu)}%
\thanks{Wing Lam is with George Mason University, Fairfax, VA 22030 USA. (e-mail: winglam@gmu.edu)}%
\thanks{Stefan Winter is with LMU Munich, Germany. (e-mail: sw@stefan-winter.net)}%
\thanks{This work has been submitted to the IEEE for possible publication. Copyright may be transferred without notice, after which this version may no longer be accessible.}%
}

\maketitle

\begin{abstract}
Flaky tests are tests that nondeterministically pass and fail in unchanged code. These tests can be detrimental to developers' productivity.
Particularly when tests run in continuous integration environments, the tests may be competing for access to limited computational resources (CPUs, memory etc.), and we hypothesize that resource (in)availability may be a significant factor in the failure rate of flaky tests.
We present the first assessment of the impact that computational resources have on flaky tests, including a total of \numProj{} projects written in Java, JavaScript and Python, and \numConfigs{} different resource configurations.
Using a rigorous statistical methodology, we determine which tests are RAFTs (Resource-Affected Flaky Tests).
We find that \percentRAFT{}\% of the flaky tests in our dataset are RAFTs, indicating that a substantial proportion of flaky-test failures can be avoided by adjusting the resources available when running tests.
We report RAFTs and configurations to avoid them to developers, and received interest to either fix the RAFTs or to improve the specifications of the projects so that tests would be run only in configurations that are unlikely to encounter RAFT failures.
Our results also have implications for researchers attempting to detect flaky tests, e.g., reducing the resources available when running tests is a cost-effective approach to detect more flaky failures.
\end{abstract}

\section{Introduction}
\label{sec:intro}




Flaky tests are tests that can pass and fail in repeated executions without changes to the test code or the code under test~\cite{Luo2014HEM}. Flaky tests are detrimental to developer's productivity.
In a continuous integration environment where developers run tests after making code changes, test failures signal to developers that their changes may have introduced a fault, which needs to be debugged and repaired so that all tests pass again.
When a flaky test fails, the developers, unaware of the flakiness at first, may be misled to debug the test failure in the recent code changes, even though the flaky test failure is unrelated to the changes and can be due to a myriad of reasons, such as dependency on specific thread interleavings, test execution orders, etc.~\cite{Luo2014HEM,Eck2019PCB,Parry2022KHMTOSEM}.
The negative effects of flaky tests have been reported as a substantial issue in many software companies, such as Apple~\cite{Kowalczyk2020NGSLM}, Ericsson~\cite{Malm2020MJ,Rehman2021R}, Facebook~\cite{ResearchFacebookTesting,Harman2018H}, Google~\cite{GoogleTestingToT,Memon2017GNDNSM,Micco2017,Ziftci2017R}, Huawei~\cite{Jiang2017LYX}, Microsoft~\cite{Herzig2015GCM,Herzig2015N,Lam2019GNST,Lam2020MST,Leesatapornwongsa22flakerepro}, and Mozilla~\cite{DeveloperTestVerification,Rahman2018R}\Space{
and triggered intensive research over the past years~\cite{Parry2022KHMTOSEM, Luo2014HEM, Eck2019PCB}, mostly pertaining to detecting and repairing flaky tests of certain types, such as implementation-dependent tests~\cite{Shi2016GLM,Zhang2021JWSMS} or order-dependent tests~\cite{Zhang2014JWMLEN, Lam2019OSMX, Shi2019LOXM, Lam2020SOZEX, Wei2021YXML}}.

This paper makes the observation that test flakiness can often be attributed to the (un)-availability of computational resources, e.g., CPU, memory, etc.
We use the term \textbf{RAFT} (Resource-Affected Flaky Test) to refer to a test that manifests flakiness under such circumstances. 
Intuitively, the unavailability of required but unspecified computational resources can lead to runtime errors that affect test execution.
For instance, if CPU resources are unavailable, either due to test execution on a weakly equipped machine or CPU contention in a multi-processing multi-tenant setting, implicit test assumptions on the latency of asynchronous operations may be violated and lead to test failures~\cite{silva2020shake,Leesatapornwongsa22flakerepro,Malm2020MJ,Lam2020MST}. Overall, the (un-)availability of resources can trigger nondeterministic behavior associated with different causes of flakiness~\cite{luo-etal-fse2014}, e.g., \textsc{Async Wait} or \textsc{Concurrency}. 


\begin{wrapfigure}{r}{0.21\textwidth}
\includegraphics[trim={16.3cm 25cm 4cm 17.5cm},clip,width=.45\textwidth]{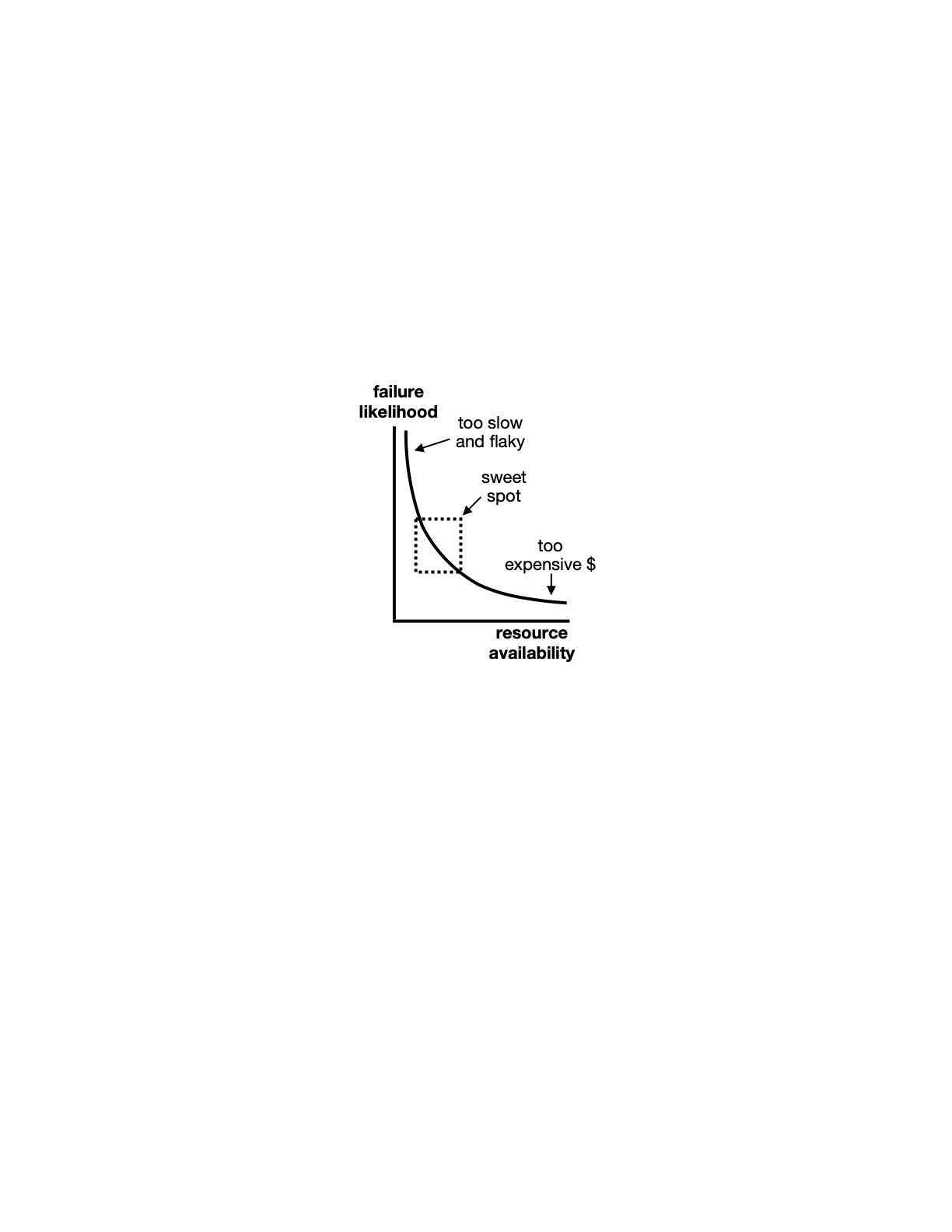}
\caption{\label{fig:raft-tradeoff}\Space{Schematic illustration of t}Trade-off between the likelihood of observing RAFT failures and resource availability\Space{ for one run of a RAFT}.}
\end{wrapfigure}

Acquiring unlimited resources is not a realistic solution to address RAFT failures as computational resources are finite, and cloud computing costs can quickly add up.
Increasing resources incessantly will eventually result in diminishing returns in terms of RAFT failure prevention relative to cost. 
Likewise, maximizing the savings in computing resources may be disruptive and unproductive due to an increase in RAFT failures. 
Figure~\ref{fig:raft-tradeoff} illustrates the trade-off between resource availability and the likelihood of a RAFT failure\Space{ during a test run}. Conceptually, the ``sweet-spot'' region in the figure represents computational resource configurations that balance cost and failure ratio.

RAFT opens an interesting perspective on the link between a controlled parameter of the execution environment---the computing resources---and the \emph{detection} and \emph{prevention} of flaky tests:
\begin{enumerate}
    \item If resources are constrained, the likelihood for observing RAFT failures increases.\Space{\footnote{We observed rare cases where RAFT is detected with the availability of more resources\Fix{cite Jon's example. websocket?}.}} This increase helps RAFT \emph{detection}, which is a prerequisite for localization and repair, in case the failure probability is very low in normal operation.
    A higher failure rate is also beneficial for debugging a flaky test.
    \item If tests are known to be sensitive to resource constraints, the resource configuration of their runtime environment can be chosen in a way to reduce failure probability and thereby \emph{prevent} RAFT from affecting developers. Future research might examine specific runtime techniques to optimize test execution. 
\end{enumerate}

To assess the effects of computational resources on flaky tests, we propose a rigorous statistical methodology to determine which flaky tests are resource-affected.
Using this methodology, we conduct a comprehensive study involving \numProj{} open-source projects written in Java, JavaScript and Python.
We structured the study in three parts:

First, we measure the prevalence of RAFT when running tests for \numTestRuns{} times on \numConfigsPhaseOne\ configurations of CPU, memory, disk, and network. 
\emph{A RAFT is a test whose failure rates differ with statistical significance under throttling and non-throttling conditions (\textsection~\ref{sec:research-questions})}. We find that nearly half of all tests found to be flaky are RAFT. 
Determining that a test is RAFT is important because it can make it easier for developers to avoid flaky failures (by providing tests with sufficient resources) and for researchers to detect flaky tests (by providing tests with fewer resources).

Second, we measure which resources have the highest impact on RAFTs.
We find that CPU availability has higher importance compared to memory and much higher importance compared to disk and network.
This finding is of high practical relevance for detection and prevention of RAFTs, as it is a parameter that can be controlled and scaled in real-world cloud computing configurations.
In our dataset of \numProj{}\Space{ Java, JavaScript, and Python} projects, we show that RAFTs are more likely to occur when the available CPU is less than 1 core and memory is less than 1GiB.
This finding is relevant for tests that can run on different hardware configurations or on shared hardware with load that varies over time, which can both lead to resource contention.
It also has important implications for debugging flaky test failures, as debugging is commonly conducted on a \emph{different} machine (e.g., a developer's laptop) than the machine on which the test failed in continuous integration.

Third, we show that scaling resources beyond certain configurations yields no statistically significant improvements for preventing RAFTs.
This finding has practical implications for reducing operational cost while mitigating RAFT.
Furthermore, we find that RAFTs that manifest nondeterministic behavior in only one of the fifteen configurations we analyzed are rare, suggesting that a small sample of configurations can be used to detect RAFT and that increasing the number of trials should suffice to increase confidence levels that a flaky test is a RAFT.
Finally, we assess the most cost-effective configurations for preventing and detecting RAFT. 

Overall, we present initial, yet strong evidence of the importance of RAFTs for regression testing. Our findings have several implications for developers (\textsection~\ref{sec:implications-developers}) and researchers (\textsection~\ref{sec:implications-researchers}) and open an avenue for further research on flaky tests.
Our dataset and scripts are publicly available under the following repository~\cite{raft-dataset}.

\section{Background and Related Work}
\label{sec:background}



Flaky tests \cite{Parry2022KHMTOSEM} have been the subject of systematic academic studies for almost a decade with numerous contributions to their detection, repair, avoidance, and tolerance at run-time.
As the root causes behind the non-determinism of flaky tests are highly diverse, so are the strategies to effectively cope with them.
Luo et al.~\cite{luo-etal-fse2014} identified \Num{10} diverse root causes for test flakiness across \Num{51} affected projects from the Apache Software Foundation and derived corresponding repair strategies from fixing commits.
Much of the following work to combat flaky tests consequently focused on individual root causes.
iDFlakies~\cite{Lam2019OSMX} and iFixFlakies~\cite{Shi2019LOXM}, for instance, have been developed as approaches for detecting and automatically repairing order-dependent flaky tests.
In their work, the authors make a terminological distinction between order-dependent and non-order-dependent flaky tests, i.e., an explicit naming of the fraction their approaches aim to address vs. the totality of flaky tests.
While other work is not making similarly dichotomous distinctions, the addressed root causes are commonly named explicitly, e.g., \emph{Assumed deterministic implementations of nondeterministic specifications}~\cite{Shi2016GLM}\MyComment{(ADINS)} or \emph{infrastructure-dependent flaky tests}~\cite{Gruber2021LKF}.
RAFTs, which are the focus of this work, adds to the collection of previously analyzed root causes of flaky tests.

Besides the extensive study of RAFTs' prevalence across a large variety of popular projects along with mitigation strategies (detection \emph{and} prevention), the focus on RAFTs conceptually distinguishes our work from a technically similar proposal by Terragni et al.~\cite{Terragni2020SF}.
While Terragni at al. also hypothesize an effect of resource unavailability on flaky test executions, their stated goal is \emph{root-causing} in the sense of deriving a flaky test's category from a number of different possible categories, some of which are not resource-related (e.g., order-dependency).
Our work, in contrast, is focusing on resource effects and explores two mitigation strategies.
For this purpose, our work leverages resource control offered by Linux control groups, which provides uniform resource access control over the entire duration of a test execution, whereas Terragni et al.'s proposal and other work (Shaker \cite{silva2020shake}) rely on dynamic load generation, with which tests compete for resources.

\subsection{Example RAFT}

\begin{figure}[t]
	\begin{lstlisting}[language=Java,basicstyle=\ttfamily\footnotesize,escapeinside={(*@}{@*)},columns=fixed,xleftmargin=.1ex]
@Test public void test() throws ScriptCPUAbuseException, ScriptException {
  NashornSandbox sandbox = 
    NashornSandboxes.create();(*@\label{fig:background:example:create}@*)
  sandbox.setMaxCPUTime(100);(*@\label{fig:background:example:setcpu}@*)
  sandbox.setMaxMemory(50 * 1024);(*@\label{fig:background:example:setmem}@*)
  ...
  ExecutorService executor = 
    Executors.newSingleThreadExecutor();
  sandbox.setExecutor(executor);
  sandbox.eval("function callMe() { return 42; };");(*@\label{fig:background:example:setfun}@*)
  Assert.assertTrue(sandbox.get("callMe") != null);(*@\label{fig:background:example:callfun}@*)
  executor.shutdown(); 
}
    \end{lstlisting}
    \precaptionspace
	\caption{Example RAFT from the \CodeIn{TestGetFunction} class in the \CodeIn{delight-nashorn-sandbox} project~\cite{GitHubDelightNashorn}.}
	\label{fig:background:example}
	\postcaptionsCodespace
\end{figure}


\sloppy
Figure~\ref{fig:background:example} shows an example of a RAFT. 
The test is from the open-source project \texttt{delight-nashorn-sandbox}~\cite{GitHubDelightNashorn} that provides a sandbox (i.e., an isolated environment) for executing JavaScript code from within Java applications.
A sandbox allows developers to limit the resources, such as maximum CPU time and memory usage that a sandbox can use, so that the sandbox will terminate if it is stuck (e.g., in an infinite loop).
In this test, Line~\ref{fig:background:example:create} first creates the sandbox that will be used.
Line~\ref{fig:background:example:setcpu} then sets the maximum amount of CPU time in milliseconds, and Line~\ref{fig:background:example:setmem} sets the maximum amount of thread memory in bytes that the JavaScript code (Line~\ref{fig:background:example:setfun}) can take to run.
When we run this test in different resource configurations, we find that the execution of the JavaScript code from Line~\ref{fig:background:example:callfun} can raise \CodeIn{Script\-CPU\-Abuse\-Exception}, \CodeIn{Script\-Memory\-Abuse\-Exception}, or raise no exceptions. 
Specifically, when we run this test on a 4\,CPU, 16\,GiB RAM machine, we find the test fails twice out of \numTestRuns{} runs -- once with each type of exception.
As we throttle the CPU and thread memory more, we see the expected number of exception increase\Space{s in 300 runs}.
For example, with 0.1\,CPU and 16\,GiB RAM, the test fails \Num{80} times, obtaining \Num{64} \CodeIn{Script\-CPU\-Abuse\-Exception}, and with 4\,CPU and 0.5\,GiB RAM, the test fails \Num{18} times, obtaining \Num{17} \CodeIn{Script\-Memory\-Abuse\-Exception}.\OurComment{from Denini: <0.1cpu, unrestricted RAM>: 80 failures: ScriptCPUAbuseException 64, ScriptMemoryAbuseException 16
<unrestricted cpu, 0.5GiB ram>: 18 failures: ScriptCPUAbuseException 1 , ScriptMemoryAbuseException 17
<0.1CPU, 1GiB ram>: 76 failures: 13 ScriptMemoryAbuseException, 63 ScriptCPUAbuseException}
Note that the test is flaky because running the Java application and the sandbox both require a nontrivial amount of CPU and RAM, and not because the throttled CPU and RAM settings exceeded what the sandbox specified as limits for the JavaScript code (lines~\ref{fig:background:example:setcpu} and~\ref{fig:background:example:setmem}), i.e., even when we increase the CPU time the sandbox can use (changing Line~\ref{fig:background:example:setcpu} to \Num{1000} instead of \Num{100}), the failure rates remain largely the same.
When we evaluate how often this test fails in machine configurations offered by Amazon Web Services (AWS), we find that the test can fail as frequently as \Num{76} times in \Num{300} runs.
More details about how often this test fails in various AWS configurations are in \textsection\ref{sec:results:rq1}. 
We report further interesting examples of flaky tests\Space{ that are and are not related to resource availability} in \textsection\ref{sec:discussion}.


\OurComment{
This flaky test fails in a normal execution (at a low rate), to see it fail more often we just need to run it in a more restricted environment of CPU and RAM.
}




\section{Methods and Objects of Analysis}
\label{sec:method-and-objects}


This section describes the projects we used (\textsection~\ref{sec:projects}), the setup of our experiments  (\textsection~\ref{sec:experimental-setup}), and the research questions we posed~(\textsection~\ref{sec:research-questions}).


\subsection{Projects}
\label{sec:projects}
Our empirical study includes projects written in three languages: Java, Python and JavaScript.
For each language, we selected projects by examining the literature to identify projects previously studied in the context of flakiness.
We included projects studied by prior work if we could build the project, run the tests, and parse the test output.
In cases where projects had missing dependencies (or other infrastructure-related failures), we spent up to three hours per-project manually debugging them, improving our tools if necessary.
For each project in our dataset, we create Docker containers that have all of the project's dependencies included, ensuring durable reproducibility and reducing the effort needed by researchers in the future to build on our results.

\subsubsection{Java}
The corpus of Java projects consists of \numProjJava{}~GitHub projects selected from two datasets: \numFlakeFlaggerDataset{}~projects from the FlakeFlagger dataset~\cite{Alshammari2021MHB} and \numLamDataset{}~projects from the Lam~\etal{}~dataset~\cite{LamETAL20ISSRE}. 
We use the same versions of each of these projects as studied in our and others' prior work.
This set of projects has been extensively studied in the context of flaky tests, and was originally built by searching GitHub for issues or commits related to flaky tests.
We excluded \Num{four} projects from the original FlakeFlagger dataset (\Num{three} of which manifested deadlocks and \Num{one} had a broken build) and excluded \Num{five} projects duplicated in the Lam~\etal{}~dataset (all of which are also included in the FlakeFlagger dataset). 

\subsubsection{Python}
The corpus of Python projects consists of \numProjPython{} projects selected from Parry \etal's recent studies~\cite{Parry22Evaluating, Parry23Empirically}.
These Python projects were selected at random from a list of projects critical to open-source infrastructure.
We first attempted to use the exact same versions of these projects that had been studied in prior work, but despite significant and generous assistance from the authors, were unable to successfully build those old versions due to missing dependencies.
We \emph{did} succeed at building the most recent revision of \Num{21} of these projects (excluding five), and created container images with those dependencies cached to ensure durable reproducibility. 
For nine of these projects, we did not observe a single flaky test during any test execution, leaving us with \numProjPython{} Python projects, that we were able to build and for which we observed at least one flaky test.

\subsubsection{JavaScript}
We used a similar methodology to select \numProjJS{} JavaScript (JS) projects, beginning by examining the projects studied in Barbosa \etal's investigation of flaky tests across programming languages (\Num{six} JS projects with at least five flaky test)~\cite{barbosa2022test}, and Yost's flaky test detection work~\cite{yostFlakyTestJSThesis} (\Num{58} JS projects).
We used our NPM-Filter infrastructure~\cite{Arteca22NPMFilter} for the building projects, running their test suites and parsing the test results.
We ran each project under NPM-Filter, and included in our corpus each project that completed within \Num{three} hours, and for which NPM-Filter could parse test results (i.e., those using the Mocha or Jest test runners).
We supplemented this set of JavaScript projects with three projects that we had previously encountered flaky tests in: \texttt{ngrok}, \texttt{IcedFrisby} and \texttt{twilio-video-app-react}.
Ultimately, this resulted in a corpus of \numProjJS{} JavaScript projects with flaky tests.

These datasets are a part of recent research on flaky test detection, which makes them ideal targets for our study and provides baselines against which our results can be compared.
Table~\ref{lbl:rq1} lists all of these projects in alphabetical order grouped by language. Our supplementary artifact~\cite{raft-dataset} contains URLs for the projects analyzed including corresponding revisions used, along with links to docker images that contain the projects packaged with all necessary dependencies to reproduce their test suites.
 

\subsection{Experimental setup}
\label{sec:experimental-setup}


The experiment consists of two phases:
\begin{description}
    \item Phase I: This phase is designed to identify the most prominent resource(s) responsible for RAFT.
    \item Phase II: This phase is designed to identify the most economically prudent real-world configurations for detection and prevention of RAFT. 
\end{description}
During both phases, the base machine consists of a virtual machine in our VMWare private cloud.
All experiments are run on virtual machines that are allocated 4 CPU cores and 16 GiB of RAM.
Within these virtual machines, we run the test suites in Docker containers, using Docker to further restrict the resources available to the test suite. 
Each experiment is implemented as a series of ``jobs,'' where each job includes the execution of one test suite under one resource-availability configuration.
The invocation of the experiments is managed using Slurm~\cite{Yoo03Slurm}, which schedules the execution of each job of each experiment on our cluster.
To prevent interference between experiments, only a single container was run at a time within any virtual machine.
It is worth noting that flaky test failures are inherently nondeterministic.
This randomness can be problematic due to its potential to skew results on a particularly lucky (or unlucky) run. 
For that reason, we run each of the \numProj{} projects on every configuration \numTestRuns{}~times and record the test failures for each run.


\begin{table}[t!]
    \setlength{\tabcolsep}{9pt} 
    \centering \caption{\label{tab:throttling-configs}Throttling configurations for Phase I. \textnormal{The highlighted row shows the default configuration (no throttling). Empty cells indicate that the value of the corresponding cell is equivalent to that of the default configuration (Baseline).}}
     \vspace{-2ex}
    \begin{tabular}{lrrrr}
    \toprule
    \# & C & M & D & N \\    
    \midrule
 \rowcolor{gray!6}   Baseline& 4 & 16 & Unrestricted & Unrestricted \\
    \cpuOnly{} & 0.1 &  &  &   \\
 \rowcolor{gray!6}    \memOnly{} &  & 0.5 &  &   \\
    \diskOnly{} &  &  & 50/100 Kbps &   \\
  \rowcolor{gray!6}   \networkOnly{} &  &  &  &  1500/512 Kbps\\
    \cpuMem & 0.1 & 0.5 &   &  \\ 
     \rowcolor{gray!6}\cpuNetwork & 0.1 &   &   &  1500/512 Kbps\\
    \memNetwork & & 0.5 &  &  1500/512 Kbps\\
     \rowcolor{gray!6}\cpuDisk & 0.1 &  & 50/100 Kbps & \\
    \memDisk &  & 0.5 & 50/100 Kbps & \\
     \rowcolor{gray!6}\diskNetwork &  &  & 50/100 Kbps & 1500/512 Kbps\\
    \cpuMemNetwork & 0.1 & 0.5 &  & 1500/512 Kbps\\
     \rowcolor{gray!6}\cpuMemDisk & 0.1 & 0.5 & 50/100 Kbps & \\ 
    \cpuDiskNetwork & 0.1 &  & 50/100 Kbps & 1500/512 Kbps\\
     \rowcolor{gray!6}\memDiskNetwork & & 0.5 & 50/100 Kbps & 1500/512 Kbps\\
    \cpuMemDiskNetwork & 0.1 & 0.5 & 50/100 Kbps & 1500/512 Kbps\\
    \bottomrule
    \end{tabular}
    \vspace{-2ex}
\end{table}

\emph{Phase I}:
To understand the impact of different resources (CPU, Memory, Disk, and Network), it is necessary to have control over them and have the ability to restrict them independently.
Table~\ref{tab:throttling-configs} shows the complete list of throttling configurations used during Phase I of our study. 
Column "\#" shows the configuration ID, column ``C'' shows the number of CPUs, column ``M'' shows the amount of memory in GiBs. 
The column ``D'' (abbreviates Disk) shows the limited rate of  IO operations per second and the throughput in kilobit per second (Kbps), respectively.
Finally, the column ``N'' (abbreviates Network) shows the network limit for download and upload in Kbps respectively. 
The options for CPU, RAM, and disk throttling are set using the Docker CLI.
The option for networking throttling is set using Wondershaper.\footnote{\url{https://github.com/magnific0/wondershaper}}


The first row on Table~\ref{tab:throttling-configs} shows the default configuration, where resources are \emph{not} throttled. 
The non-default configurations \cpuOnly{}-\cpuMemDiskNetwork{} modify the values assigned to one or more resources. 
We chose very small values to assign to each configuration option (i.e., resource) with the goal of running the tests under ``limit'' conditions.
The configurations \cpuOnly{}-\networkOnly{} throttle only a single resource at a time, and are useful to understand the impact of individual resources on flaky failures. 
We also consider all combinations of those parameters, which helps us to answer whether certain tests are RAFTs only when multiple resources are constrained simultaneously.


\emph{Phase II}:
In order to identify the most cost effective real-world cloud computing configurations for detection and prevention of RAFTs, we examine resource configurations that more closely match those available by major cloud providers. 
Cloud providers offer flexible pricing for on-demand containers-as-a-service, e.g. AWS Fargate~\cite{aws-fargate}, Google Kubernetes Engine~\cite{google-kubernetes-engine}, and Azure Kubernetes Service~\cite{azure-kubernetes-service}.
These services are priced by CPU and memory specifications, and provide ``standard'' disk and network access services.
During Phase I, we conclude that CPU has a greater influence on test flakiness compared to memory, disk and network.
As a result, in Phase II, we consider configurations with a different number of CPUs, assigning the lowest and the highest memory options available on AWS for each configuration.
No restrictions are imposed on the disk and network during this phase.
As in Phase I, we ran each test suite \numTestRuns{} times.
Table~\ref{tab:aws-configs} shows the complete list of configurations used during Phase II of our study.
Column ``\#'' shows the configuration id, column "CPU" shows the number of CPUs, column ``Mem(GiB)'' shows the amount of memory in GiBs.
Column ``Cost (\$/hr)" shows the AWS computing cost per hour of a given CPU and Memory configuration on AWS Fargate~\cite{aws-fargate} serverless compute engine. In AWS Fargate, developers submit Dockers images to the engine and pay for compute resources when used.
We consider \Num{12} combinations of CPU, ranging from 0.1 to 4, and of memory, ranging from 1~GiB to 16~GiB.
The AWS computing cost (measured in USD per hour) varies with the quality of the service and the configuration requested~\cite{aws-fargate-pricing}. For a given configuration, the cost of the ``spot'' service is lower compared to the cost of the ``on-demand'' service.
Whereas the ``on-demand'' service provides guaranteed availability of the container, a ``spot'' container may be interrupted and canceled by the service provider to shed their load during peak usage times.
However, the significant savings may make it attractive for running test suites in CI, where a canceled test suite can be restarted on another container.

\begin{table}[t!]
    \setlength{\tabcolsep}{8pt} 
    \centering
     \caption{\label{tab:aws-configs}AWS configurations sorted by cost~\cite{aws-fargate-pricing}. \textnormal{Disk and Network are unrestricted}.}
    \vspace{-2ex}
    \begin{tabular}{rrrrr}
    \toprule
    \multirow{2}{*}{\#} & \multirow{2}{*}{CPU} & \multirow{2}{*}{Mem (GiB)} & \multicolumn{2}{c}{Cost (USD/hour)} \\    
     & & & \multicolumn{1}{c}{spot} & \multicolumn{1}{c}{on-demand}\\
    \midrule
    1 & 0.1 & 1 & 0.002548 & 0.008493\\ 
     \rowcolor{gray!6}2 & 0.1 & 2 & 0.003881 & 0.012938\\ 
    3 & 0.25 & 2 & 0.005703 & 0.019010\\ 
     \rowcolor{gray!6}4 & 0.5 & 2 & 0.008739 & 0.029130\\ 
    5 & 0.5 & 4 & 0.011406 & 0.038020\\ 
    \rowcolor{gray!6} 6 & 1 & 4 & 0.017478 & 0.058260\\ 
    7 & 1 & 8 & 0.022812 & 0.076040\\ 
    \rowcolor{gray!6} 8 & 2 & 4 & 0.029622 & 0.098740\\ 
    9 & 2 & 8 & 0.034956 & 0.116520\\ 
    \rowcolor{gray!6} 10 & 2 & 16 & 0.045624 & 0.152080\\
    11 & 4 & 8 & 0.059244 & 0.197480\\ 
   \rowcolor{gray!6}  12 & 4 & 16 & 0.069912 & 0.233040\\
    \bottomrule
    \end{tabular}
\vspace{-3.5ex}    
\end{table}

\subsection{Research questions and methodology
}
\label{sec:research-questions}

We aim to answer the following key research questions:

\subsection*{\textbf{RQ1.} \rqone{}} 
\noindent
\textbf{Rationale.} This question is important to justify further investigation on \emph{Resource-Affected Flaky Tests} (RAFTs).
If we find that none of the flakiness can be attributed to resource starvation, then further investigation is meaningless.
To understand the prevalence of RAFTs in flaky test failures, we aim to answer two key questions:
\begin{description}
    \item \textbf{RQ1.1.} \rqoneone{}
    \item \textbf{RQ1.2.} \rqonetwo{}
\end{description}
RQ1.1 aims to distinguish RAFT failures from other kinds of failure to establish their prevalence.
RQ1.2 aims to quantify the effect of resource starvation on RAFT failures to establish how likely these failures are under resource throttling.



\noindent
\textbf{Methodology.} To answer RQ1, we consider the data for the following attributes for each project: (i)~the number of flaky tests identified under full resource availability, (ii)~the number of flaky tests identified under all test execution configurations combined (Table~\ref{tab:throttling-configs}), and (iii)~the number of test failures which can be considered RAFTs.
Since a definition for RAFTs does not exist, we present the first quantifiable definition of RAFTs.

\noindent\textbf{Definition:} A \emph{Resource-Affected Flaky Test} (RAFT) is a flaky test that has a statistically different failure rate when resources are constrained compared to an unconstrained test execution.
We use Pearson's chi-squared test to determine whether the failure rate is statistically different, accepting that difference as significant only at a level of $p<0.05$.
To reduce the false discovery rate, we use the  Benjamini-Hochberg procedure to adjust p-values.
For a given throttling configuration, a test can be a RAFT only if it passed at least once under that same configuration.

We use the three attributes above to answer RQ1.1. 
To answer RQ1.2, we consider the increase in failure rates for every unique test failure. Such increase is defined as the ratio $f_i/max(f_1,1)$, where $f_i$ is the number of failures under the most failure-inducing throttling configuration and $f_1$ is the number of failures under the configuration with no resource throttling.
The ratios are then grouped by different levels of increase in failure rate.

\subsection*{\textbf{RQ2.} \rqtwo{}}
\noindent
\textbf{Rationale}. This question is important to justify further investigation of the relationship between resource availability and test flakiness. 
If we find that the relationship is weak, then further investigation is meaningless.
To understand the effect of machine resources on flaky test failures, we aim to answer two questions: 
\begin{description}
    \item \textbf{RQ2.1.} \rqtwoone{}
    \item \textbf{RQ2.2.} \rqtwotwo{}
\end{description}
RQ2.1 aims to study the effect that throttling individual resources has on flaky test failures. RQ2.2 aims to study the effect that throttling combinations of resources has on flaky test failures and whether this produces results that are significantly different to throttling individual resources.
It is important to explore each resource independently and in combinations to understand their impact.
The resources or combinations with the most impact can then be chosen for further analysis.

\noindent
\textbf{Methodology.} To answer RQ2, we compare the number of RAFTs detected under each throttling configuration shown in Table~\ref{tab:throttling-configs}.
Each throttling configuration limits availability of one to four resources.
We compare the number of RAFTs detected by each configuration, analyzing the configurations that detect each RAFT.

\subsection*{\textbf{RQ3.} \rqthree{}} 

\noindent
\textbf{Rationale}. In a typical usage of continuous integration, developers want to simultaneously~(1)~avoid flaky tests to reliably determine whether a bug is present in code when observing test failures and to~(2)~run tests efficiently, i.e., maximize test runs per amount of money. This question focuses on this scenario. More precisely, it investigates which resource configurations give the lowest flaky test disruption per amount of money for given a project.

\noindent
\textbf{Methodology.}
To answer RQ1 and RQ2, we examined the total RAFTs detected across all \numTestRuns{} test suite invocations.
To answer RQ3, we study instead the number of test suite invocations (i.e., builds) that have at least one flaky-test failure.
We consider two metrics  for every resource configuration listed in Table~\ref{tab:aws-configs}: (i) the number of builds with test failures across all test runs (as a proxy for reliability) and (ii) the price per-test suite run.
We calculate the price per-test suite run by multiplying the average time to run the project's test suite (as reported by the build system) by the ``on-demand'' AWS Fargate cost shown in Table~\ref{tab:aws-configs}.
The configurations with the least number of build failures are considered the most reliable. However, there may be other configurations which have slightly higher rate of build failures but are more cost effective. 
It is worth noting that configurations with a lower hourly rate can take longer to complete due to limited resources, resulting in a higher cost for each build compared to an expensive but fast configuration.
For every configuration in Table~\ref{tab:aws-configs}, we consider the number of projects for which it had the best price, best reliability, or both.



\subsection*{\textbf{RQ4.} \rqfour{}} 


\noindent
\textbf{Rationale.} In another use case, developers may want to~(1)~detect flaky tests in advance (i.e., before observing potentially spurious failures during regression runs) and~(2)~run tests efficiently. This question focuses on this scenario. More precisely, it investigates which configurations maximize the ability of test runs to detect flaky tests while keeping costs at a minimum.


\noindent
\textbf{Methodology.} To answer RQ4, we consider two metrics for every resource configuration listed in Table~\ref{tab:aws-configs}: (i) the number of flaky test failures (as a proxy for reliability of detection) and (ii) the price per run. 
As with RQ3, we calculate the price per-run by multiplying the average time to run the project's test suite (as reported by the build system) by the ``on-demand'' AWS Fargate cost shown in Table~\ref{tab:aws-configs}.
Intuitively, cheaper configurations are more likely to detect flaky test failures, but they may not be the most cost effective due to slower execution times.
Furthermore, some configurations may be entirely unusable for some projects - for example, when a project's tests require some minimum amount of memory to run at all.
Hence, it is necessary to consider computing cost for this analysis.
For every configuration in \ref{tab:aws-configs}, we consider the number of projects for which it had the best price, best detection, or both.

\begin{table*}
\setlength{\tabcolsep}{1pt}
\centering
\caption{\label{lbl:rq1}For each project with flaky tests, we report the baseline number of flaky tests identified without resource throttling (Baseline Flaky) and the number of flaky tests identified as RAFT under each throttling condition. 
\textnormal{Throttling conditions are identified by the resources throttled, i.e., CPU\cpuOnly, Memory\memOnly, Disk\diskOnly, Network\networkOnly, and combinations thereof. ``Total, All Runs'' summarises all flaky tests (and RAFT) detected \emph{including} the Phase II (AWS) configurations.
Some projects resulted in catastrophic failures under certain configurations that are indicated as "-".}
}
\vspace{-2ex}
\begin{tabular}{llr|rrrr|rrrrrr|rrrr|r|rr}
\toprule
&&\multirow{3}{*}{\parbox{1cm}{\centering Baseline Flaky}} & \multicolumn{15}{c|}{Flaky Tests Identified as RAFT Under Throttling Conditions} & \multicolumn{2}{c}{Total, All Runs}\\
\cmidrule(lr){4-18} \cmidrule(l){19-20}
Language & Project & & \cpuOnly & \memOnly & \diskOnly & \networkOnly & \cpuMem & \cpuNetwork & \memNetwork & \cpuDisk & \memDisk & \diskNetwork & \cpuMemNetwork & \cpuMemDisk & \cpuDiskNetwork & \memDiskNetwork & \cpuMemDiskNetwork & Flaky & RAFTs\\\midrule
\multirow{31}{*}{Java} & \cellcolor{gray!6}{assertj-core} & \cellcolor{gray!6}{1} & \cellcolor{gray!6}{\textbf{1}} & \cellcolor{gray!6}{-} & \cellcolor{gray!6}{0} & \cellcolor{gray!6}{0} & \cellcolor{gray!6}{\textbf{1}} & \cellcolor{gray!6}{\textbf{1}} & \cellcolor{gray!6}{0} & \cellcolor{gray!6}{\textbf{1}} & \cellcolor{gray!6}{-} & \cellcolor{gray!6}{0} & \cellcolor{gray!6}{-} & \cellcolor{gray!6}{\textbf{1}} & \cellcolor{gray!6}{\textbf{1}} & \cellcolor{gray!6}{0} & \cellcolor{gray!6}{\textbf{1}} & \cellcolor{gray!6}{3} & \cellcolor{gray!6}{1}\\
 & carbon-apimgt & 0 & 0 & 0 & 0 & 0 & 0 & \textbf{1} & 0 & 0 & 0 & 0 & 0 & 0 & 0 & 0 & 0 & 1 & 1\\
& \cellcolor{gray!6}{commons-exec} & \cellcolor{gray!6}{0} & \cellcolor{gray!6}{\textbf{1}} & \cellcolor{gray!6}{0} & \cellcolor{gray!6}{0} & \cellcolor{gray!6}{0} & \cellcolor{gray!6}{\textbf{1}} & \cellcolor{gray!6}{\textbf{1}} & \cellcolor{gray!6}{0} & \cellcolor{gray!6}{\textbf{1}} & \cellcolor{gray!6}{0} & \cellcolor{gray!6}{0} & \cellcolor{gray!6}{\textbf{1}} & \cellcolor{gray!6}{\textbf{1}} & \cellcolor{gray!6}{\textbf{1}} & \cellcolor{gray!6}{0} & \cellcolor{gray!6}{\textbf{1}} & \cellcolor{gray!6}{3} & \cellcolor{gray!6}{1}\\
 & db-scheduler & 5 & 2 & 0 & 0 & 0 & 2 & \textbf{6} & 0 & 2 & 0 & 0 & \textbf{6} & 2 & 2 & 0 & \textbf{6} & 7 & 6\\
& \cellcolor{gray!6}{delight-nashorn-sandbox} & \cellcolor{gray!6}{1} & \cellcolor{gray!6}{17} & \cellcolor{gray!6}{1} & \cellcolor{gray!6}{0} & \cellcolor{gray!6}{0} & \cellcolor{gray!6}{\textbf{24}} & \cellcolor{gray!6}{20} & \cellcolor{gray!6}{0} & \cellcolor{gray!6}{16} & \cellcolor{gray!6}{0} & \cellcolor{gray!6}{0} & \cellcolor{gray!6}{\textbf{24}} & \cellcolor{gray!6}{23} & \cellcolor{gray!6}{19} & \cellcolor{gray!6}{0} & \cellcolor{gray!6}{23} & \cellcolor{gray!6}{27} & \cellcolor{gray!6}{24}\\
 & elastic-job-lite & 0 & 0 & - & 0 & 0 & - & 0 & 0 & 0 & 0 & \textbf{1} & - & - & 0 & - & - & 1 & 1\\
& \cellcolor{gray!6}{esper} & \cellcolor{gray!6}{1} & \cellcolor{gray!6}{0} & \cellcolor{gray!6}{0} & \cellcolor{gray!6}{0} & \cellcolor{gray!6}{0} & \cellcolor{gray!6}{0} & \cellcolor{gray!6}{0} & \cellcolor{gray!6}{0} & \cellcolor{gray!6}{0} & \cellcolor{gray!6}{0} & \cellcolor{gray!6}{0} & \cellcolor{gray!6}{0} & \cellcolor{gray!6}{0} & \cellcolor{gray!6}{0} & \cellcolor{gray!6}{0} & \cellcolor{gray!6}{0} & \cellcolor{gray!6}{1} & \cellcolor{gray!6}{0}\\
 & excelastic & 1 & \textbf{1} & 0 & 0 & 0 & \textbf{1} & \textbf{1} & 0 & \textbf{1} & 0 & 0 & \textbf{1} & \textbf{1} & \textbf{1} & 0 & \textbf{1} & 2 & 1\\
& \cellcolor{gray!6}{fastjson} & \cellcolor{gray!6}{0} & \cellcolor{gray!6}{0} & \cellcolor{gray!6}{-} & \cellcolor{gray!6}{0} & \cellcolor{gray!6}{0} & \cellcolor{gray!6}{\textbf{2}} & \cellcolor{gray!6}{0} & \cellcolor{gray!6}{-} & \cellcolor{gray!6}{0} & \cellcolor{gray!6}{-} & \cellcolor{gray!6}{0} & \cellcolor{gray!6}{\textbf{2}} & \cellcolor{gray!6}{\textbf{2}} & \cellcolor{gray!6}{1} & \cellcolor{gray!6}{-} & \cellcolor{gray!6}{\textbf{2}} & \cellcolor{gray!6}{3} & \cellcolor{gray!6}{2}\\
 & fluent-logger-java & 0 & 1 & 0 & 0 & 0 & \textbf{2} & 1 & 0 & 1 & 0 & 0 & \textbf{2} & \textbf{2} & 1 & 0 & \textbf{2} & 5 & 2\\
& \cellcolor{gray!6}{handlebars.java} & \cellcolor{gray!6}{1} & \cellcolor{gray!6}{0} & \cellcolor{gray!6}{0} & \cellcolor{gray!6}{0} & \cellcolor{gray!6}{0} & \cellcolor{gray!6}{0} & \cellcolor{gray!6}{0} & \cellcolor{gray!6}{0} & \cellcolor{gray!6}{0} & \cellcolor{gray!6}{0} & \cellcolor{gray!6}{0} & \cellcolor{gray!6}{0} & \cellcolor{gray!6}{0} & \cellcolor{gray!6}{0} & \cellcolor{gray!6}{0} & \cellcolor{gray!6}{0} & \cellcolor{gray!6}{1} & \cellcolor{gray!6}{0}\\
 & hector & 2 & 0 & 0 & 0 & 0 & 0 & - & 0 & - & 0 & 0 & - & 0 & - & 0 & - & 2 & 0\\
& \cellcolor{gray!6}{http-request} & \cellcolor{gray!6}{0} & \cellcolor{gray!6}{0} & \cellcolor{gray!6}{0} & \cellcolor{gray!6}{0} & \cellcolor{gray!6}{0} & \cellcolor{gray!6}{0} & \cellcolor{gray!6}{0} & \cellcolor{gray!6}{0} & \cellcolor{gray!6}{0} & \cellcolor{gray!6}{0} & \cellcolor{gray!6}{0} & \cellcolor{gray!6}{0} & \cellcolor{gray!6}{0} & \cellcolor{gray!6}{0} & \cellcolor{gray!6}{0} & \cellcolor{gray!6}{0} & \cellcolor{gray!6}{6} & \cellcolor{gray!6}{0}\\
 & httpcore & 4 & 1 & 0 & 0 & 0 & 1 & 1 & 0 & 1 & 0 & 0 & \textbf{2} & 1 & 1 & 0 & 1 & 22 & 2\\
& \cellcolor{gray!6}{hutool} & \cellcolor{gray!6}{1} & \cellcolor{gray!6}{0} & \cellcolor{gray!6}{0} & \cellcolor{gray!6}{0} & \cellcolor{gray!6}{0} & \cellcolor{gray!6}{0} & \cellcolor{gray!6}{0} & \cellcolor{gray!6}{0} & \cellcolor{gray!6}{0} & \cellcolor{gray!6}{0} & \cellcolor{gray!6}{0} & \cellcolor{gray!6}{0} & \cellcolor{gray!6}{0} & \cellcolor{gray!6}{0} & \cellcolor{gray!6}{0} & \cellcolor{gray!6}{0} & \cellcolor{gray!6}{1} & \cellcolor{gray!6}{1}\\
 & incubator-dubbo & 3 & 23 & - & 0 & 0 & - & \textbf{25} & - & \textbf{25} & - & 0 & - & - & 22 & - & - & 55 & 27\\
& \cellcolor{gray!6}{java-websocket} & \cellcolor{gray!6}{22} & \cellcolor{gray!6}{16} & \cellcolor{gray!6}{7} & \cellcolor{gray!6}{0} & \cellcolor{gray!6}{1} & \cellcolor{gray!6}{22} & \cellcolor{gray!6}{20} & \cellcolor{gray!6}{5} & \cellcolor{gray!6}{18} & \cellcolor{gray!6}{8} & \cellcolor{gray!6}{0} & \cellcolor{gray!6}{\textbf{24}} & \cellcolor{gray!6}{\textbf{24}} & \cellcolor{gray!6}{19} & \cellcolor{gray!6}{8} & \cellcolor{gray!6}{22} & \cellcolor{gray!6}{37} & \cellcolor{gray!6}{32}\\
 & logback & 6 & 7 & 0 & 0 & 0 & 7 & 6 & 0 & 8 & 0 & 0 & \textbf{9} & - & 6 & 0 & 8 & 28 & 12\\
& \cellcolor{gray!6}{luwak} & \cellcolor{gray!6}{0} & \cellcolor{gray!6}{0} & \cellcolor{gray!6}{0} & \cellcolor{gray!6}{0} & \cellcolor{gray!6}{0} & \cellcolor{gray!6}{\textbf{1}} & \cellcolor{gray!6}{\textbf{1}} & \cellcolor{gray!6}{0} & \cellcolor{gray!6}{\textbf{1}} & \cellcolor{gray!6}{0} & \cellcolor{gray!6}{0} & \cellcolor{gray!6}{0} & \cellcolor{gray!6}{0} & \cellcolor{gray!6}{\textbf{1}} & \cellcolor{gray!6}{0} & \cellcolor{gray!6}{0} & \cellcolor{gray!6}{4} & \cellcolor{gray!6}{2}\\
 & ninja & 3 & 0 & 0 & 0 & 0 & 0 & 0 & 0 & 0 & 0 & 0 & 0 & 0 & 0 & 0 & 0 & 3 & 0\\
& \cellcolor{gray!6}{noxy} & \cellcolor{gray!6}{0} & \cellcolor{gray!6}{0} & \cellcolor{gray!6}{0} & \cellcolor{gray!6}{0} & \cellcolor{gray!6}{0} & \cellcolor{gray!6}{0} & \cellcolor{gray!6}{0} & \cellcolor{gray!6}{0} & \cellcolor{gray!6}{0} & \cellcolor{gray!6}{0} & \cellcolor{gray!6}{0} & \cellcolor{gray!6}{0} & \cellcolor{gray!6}{0} & \cellcolor{gray!6}{0} & \cellcolor{gray!6}{0} & \cellcolor{gray!6}{0} & \cellcolor{gray!6}{1} & \cellcolor{gray!6}{0}\\
 & orbit & 2 & \textbf{2} & 0 & 0 & 0 & \textbf{2} & \textbf{2} & 0 & \textbf{2} & 0 & 0 & \textbf{2} & \textbf{2} & \textbf{2} & 0 & \textbf{2} & 6 & 2\\
& \cellcolor{gray!6}{oryx} & \cellcolor{gray!6}{1} & \cellcolor{gray!6}{0} & \cellcolor{gray!6}{0} & \cellcolor{gray!6}{0} & \cellcolor{gray!6}{0} & \cellcolor{gray!6}{0} & \cellcolor{gray!6}{0} & \cellcolor{gray!6}{0} & \cellcolor{gray!6}{0} & \cellcolor{gray!6}{0} & \cellcolor{gray!6}{0} & \cellcolor{gray!6}{0} & \cellcolor{gray!6}{0} & \cellcolor{gray!6}{0} & \cellcolor{gray!6}{0} & \cellcolor{gray!6}{0} & \cellcolor{gray!6}{2} & \cellcolor{gray!6}{0}\\
 & riptide & 0 & 1 & \textbf{2} & 0 & 0 & 1 & 1 & 0 & 1 & 0 & 0 & 1 & 1 & 1 & 0 & 1 & 5 & 3\\
& \cellcolor{gray!6}{rxjava2-extras} & \cellcolor{gray!6}{2} & \cellcolor{gray!6}{0} & \cellcolor{gray!6}{0} & \cellcolor{gray!6}{0} & \cellcolor{gray!6}{0} & \cellcolor{gray!6}{0} & \cellcolor{gray!6}{0} & \cellcolor{gray!6}{0} & \cellcolor{gray!6}{0} & \cellcolor{gray!6}{0} & \cellcolor{gray!6}{0} & \cellcolor{gray!6}{0} & \cellcolor{gray!6}{0} & \cellcolor{gray!6}{0} & \cellcolor{gray!6}{0} & \cellcolor{gray!6}{0} & \cellcolor{gray!6}{7} & \cellcolor{gray!6}{1}\\
 & spring-boot & 0 & 0 & - & 0 & 0 & - & 0 & - & 0 & - & 0 & - & - & 0 & - & - & 3 & 0\\
& \cellcolor{gray!6}{timely} & \cellcolor{gray!6}{4} & \cellcolor{gray!6}{0} & \cellcolor{gray!6}{0} & \cellcolor{gray!6}{0} & \cellcolor{gray!6}{0} & \cellcolor{gray!6}{0} & \cellcolor{gray!6}{0} & \cellcolor{gray!6}{0} & \cellcolor{gray!6}{0} & \cellcolor{gray!6}{0} & \cellcolor{gray!6}{0} & \cellcolor{gray!6}{0} & \cellcolor{gray!6}{0} & \cellcolor{gray!6}{0} & \cellcolor{gray!6}{0} & \cellcolor{gray!6}{0} & \cellcolor{gray!6}{4} & \cellcolor{gray!6}{0}\\
 & wro4j & 7 & \textbf{5} & - & 0 & 0 & - & \textbf{5} & - & \textbf{5} & - & 0 & - & - & \textbf{5} & - & - & 13 & 6\\
& \cellcolor{gray!6}{yawp} & \cellcolor{gray!6}{1} & \cellcolor{gray!6}{\textbf{1}} & \cellcolor{gray!6}{0} & \cellcolor{gray!6}{0} & \cellcolor{gray!6}{0} & \cellcolor{gray!6}{\textbf{1}} & \cellcolor{gray!6}{\textbf{1}} & \cellcolor{gray!6}{0} & \cellcolor{gray!6}{\textbf{1}} & \cellcolor{gray!6}{0} & \cellcolor{gray!6}{0} & \cellcolor{gray!6}{\textbf{1}} & \cellcolor{gray!6}{\textbf{1}} & \cellcolor{gray!6}{\textbf{1}} & \cellcolor{gray!6}{0} & \cellcolor{gray!6}{\textbf{1}} & \cellcolor{gray!6}{1} & \cellcolor{gray!6}{1}\\
 & zxing & 2 & 0 & 0 & 0 & 0 & 0 & 0 & 0 & 0 & 0 & 0 & - & 0 & 0 & 0 & 0 & 2 & 0\\
  \cmidrule{2-20}
& \cellcolor{gray!6}{30 Projects Total} & \cellcolor{gray!6}{70} & \cellcolor{gray!6}{79} & \cellcolor{gray!6}{10} & \cellcolor{gray!6}{0} & \cellcolor{gray!6}{1} & \cellcolor{gray!6}{68} & \cellcolor{gray!6}{\textbf{93}} & \cellcolor{gray!6}{5} & \cellcolor{gray!6}{84} & \cellcolor{gray!6}{8} & \cellcolor{gray!6}{1} & \cellcolor{gray!6}{75} & \cellcolor{gray!6}{61} & \cellcolor{gray!6}{84} & \cellcolor{gray!6}{8} & \cellcolor{gray!6}{71} & \cellcolor{gray!6}{256} & \cellcolor{gray!6}{128}\\
\midrule
\multirow{11}{*}{JavaScript} & apollo-client-devtools & 0 & 11 & - & 0 & 0 & - & 8 & - & 10 & - & 0 & - & - & \textbf{13} & - & - & 21 & 13\\
& \cellcolor{gray!6}{IcedFrisby} & \cellcolor{gray!6}{1} & \cellcolor{gray!6}{0} & \cellcolor{gray!6}{0} & \cellcolor{gray!6}{0} & \cellcolor{gray!6}{0} & \cellcolor{gray!6}{0} & \cellcolor{gray!6}{9} & \cellcolor{gray!6}{\textbf{10}} & \cellcolor{gray!6}{9} & \cellcolor{gray!6}{\textbf{10}} & \cellcolor{gray!6}{9} & \cellcolor{gray!6}{\textbf{10}} & \cellcolor{gray!6}{0} & \cellcolor{gray!6}{0} & \cellcolor{gray!6}{\textbf{10}} & \cellcolor{gray!6}{9} & \cellcolor{gray!6}{16} & \cellcolor{gray!6}{11}\\
 & javascript-action & 0 & 0 & 0 & 0 & 0 & 0 & 0 & 0 & 0 & 0 & 0 & 0 & 0 & 0 & 0 & 0 & 1 & 1\\
& \cellcolor{gray!6}{ngrok} & \cellcolor{gray!6}{2} & \cellcolor{gray!6}{0} & \cellcolor{gray!6}{0} & \cellcolor{gray!6}{0} & \cellcolor{gray!6}{0} & \cellcolor{gray!6}{0} & \cellcolor{gray!6}{0} & \cellcolor{gray!6}{0} & \cellcolor{gray!6}{1} & \cellcolor{gray!6}{0} & \cellcolor{gray!6}{0} & \cellcolor{gray!6}{0} & \cellcolor{gray!6}{1} & \cellcolor{gray!6}{\textbf{2}} & \cellcolor{gray!6}{0} & \cellcolor{gray!6}{1} & \cellcolor{gray!6}{8} & \cellcolor{gray!6}{2}\\
 & preset-modules & 0 & \textbf{2} & 0 & 0 & 0 & 1 & 1 & 0 & \textbf{2} & 0 & 0 & 1 & 1 & 1 & 0 & 1 & 2 & 2\\
& \cellcolor{gray!6}{react-datetime} & \cellcolor{gray!6}{0} & \cellcolor{gray!6}{\textbf{1}} & \cellcolor{gray!6}{0} & \cellcolor{gray!6}{0} & \cellcolor{gray!6}{0} & \cellcolor{gray!6}{\textbf{1}} & \cellcolor{gray!6}{\textbf{1}} & \cellcolor{gray!6}{0} & \cellcolor{gray!6}{\textbf{1}} & \cellcolor{gray!6}{0} & \cellcolor{gray!6}{0} & \cellcolor{gray!6}{\textbf{1}} & \cellcolor{gray!6}{\textbf{1}} & \cellcolor{gray!6}{\textbf{1}} & \cellcolor{gray!6}{0} & \cellcolor{gray!6}{\textbf{1}} & \cellcolor{gray!6}{2} & \cellcolor{gray!6}{2}\\
 & react-native & 1 & 3 & 0 & 0 & 0 & 3 & 3 & 0 & 3 & 0 & 0 & - & 3 & \textbf{4} & 0 & 3 & 17 & 10\\
& \cellcolor{gray!6}{shields} & \cellcolor{gray!6}{0} & \cellcolor{gray!6}{\textbf{5}} & \cellcolor{gray!6}{0} & \cellcolor{gray!6}{0} & \cellcolor{gray!6}{0} & \cellcolor{gray!6}{\textbf{5}} & \cellcolor{gray!6}{\textbf{5}} & \cellcolor{gray!6}{0} & \cellcolor{gray!6}{\textbf{5}} & \cellcolor{gray!6}{0} & \cellcolor{gray!6}{0} & \cellcolor{gray!6}{\textbf{5}} & \cellcolor{gray!6}{\textbf{5}} & \cellcolor{gray!6}{\textbf{5}} & \cellcolor{gray!6}{0} & \cellcolor{gray!6}{\textbf{5}} & \cellcolor{gray!6}{6} & \cellcolor{gray!6}{6}\\
 & tippyjs-react & 0 & 0 & 0 & 0 & 0 & 0 & 0 & 0 & 0 & 0 & 0 & 0 & \textbf{1} & \textbf{1} & 0 & 0 & 1 & 1\\
& \cellcolor{gray!6}{twilio-video-app-react} & \cellcolor{gray!6}{0} & \cellcolor{gray!6}{6} & \cellcolor{gray!6}{-} & \cellcolor{gray!6}{0} & \cellcolor{gray!6}{0} & \cellcolor{gray!6}{-} & \cellcolor{gray!6}{6} & \cellcolor{gray!6}{-} & \cellcolor{gray!6}{\textbf{8}} & \cellcolor{gray!6}{-} & \cellcolor{gray!6}{0} & \cellcolor{gray!6}{-} & \cellcolor{gray!6}{-} & \cellcolor{gray!6}{\textbf{8}} & \cellcolor{gray!6}{-} & \cellcolor{gray!6}{-} & \cellcolor{gray!6}{18} & \cellcolor{gray!6}{8}\\
 \cmidrule{2-20}
 & 10 Projects Total & 4 & 28 & 0 & 0 & 0 & 10 & 33 & 10 & \textbf{39} & 10 & 9 & 17 & 12 & 35 & 10 & 20 & 92 & 56\\
 \midrule
\multirow{13}{*}{Python} & \cellcolor{gray!6}{celery} & \cellcolor{gray!6}{0} & \cellcolor{gray!6}{0} & \cellcolor{gray!6}{0} & \cellcolor{gray!6}{0} & \cellcolor{gray!6}{0} & \cellcolor{gray!6}{0} & \cellcolor{gray!6}{0} & \cellcolor{gray!6}{0} & \cellcolor{gray!6}{0} & \cellcolor{gray!6}{0} & \cellcolor{gray!6}{0} & \cellcolor{gray!6}{0} & \cellcolor{gray!6}{0} & \cellcolor{gray!6}{0} & \cellcolor{gray!6}{0} & \cellcolor{gray!6}{0} & \cellcolor{gray!6}{1} & \cellcolor{gray!6}{0}\\
 & conan & 0 & 0 & 0 & \textbf{2} & 0 & 0 & 0 & 0 & \textbf{2} & \textbf{2} & \textbf{2} & 0 & \textbf{2} & \textbf{2} & \textbf{2} & \textbf{2} & 6 & 2\\
& \cellcolor{gray!6}{electrum} & \cellcolor{gray!6}{0} & \cellcolor{gray!6}{\textbf{1}} & \cellcolor{gray!6}{0} & \cellcolor{gray!6}{0} & \cellcolor{gray!6}{0} & \cellcolor{gray!6}{\textbf{1}} & \cellcolor{gray!6}{0} & \cellcolor{gray!6}{0} & \cellcolor{gray!6}{\textbf{1}} & \cellcolor{gray!6}{0} & \cellcolor{gray!6}{0} & \cellcolor{gray!6}{\textbf{1}} & \cellcolor{gray!6}{\textbf{1}} & \cellcolor{gray!6}{\textbf{1}} & \cellcolor{gray!6}{0} & \cellcolor{gray!6}{\textbf{1}} & \cellcolor{gray!6}{1} & \cellcolor{gray!6}{1}\\
 & fonttools & 0 & 0 & 0 & 0 & 0 & 0 & 0 & 0 & 0 & 0 & 0 & 0 & 0 & 0 & 0 & 0 & 1 & 0\\
& \cellcolor{gray!6}{ipython} & \cellcolor{gray!6}{0} & \cellcolor{gray!6}{0} & \cellcolor{gray!6}{0} & \cellcolor{gray!6}{0} & \cellcolor{gray!6}{0} & \cellcolor{gray!6}{0} & \cellcolor{gray!6}{0} & \cellcolor{gray!6}{0} & \cellcolor{gray!6}{\textbf{3}} & \cellcolor{gray!6}{0} & \cellcolor{gray!6}{0} & \cellcolor{gray!6}{0} & \cellcolor{gray!6}{\textbf{3}} & \cellcolor{gray!6}{\textbf{3}} & \cellcolor{gray!6}{0} & \cellcolor{gray!6}{\textbf{3}} & \cellcolor{gray!6}{4} & \cellcolor{gray!6}{3}\\
 & loguru & 0 & 0 & 0 & 0 & 0 & 0 & 0 & 0 & 0 & 0 & 0 & 0 & 0 & 0 & 0 & 0 & 42 & 0\\
& \cellcolor{gray!6}{mitmproxy} & \cellcolor{gray!6}{0} & \cellcolor{gray!6}{0} & \cellcolor{gray!6}{0} & \cellcolor{gray!6}{0} & \cellcolor{gray!6}{0} & \cellcolor{gray!6}{0} & \cellcolor{gray!6}{0} & \cellcolor{gray!6}{0} & \cellcolor{gray!6}{0} & \cellcolor{gray!6}{0} & \cellcolor{gray!6}{0} & \cellcolor{gray!6}{0} & \cellcolor{gray!6}{0} & \cellcolor{gray!6}{0} & \cellcolor{gray!6}{0} & \cellcolor{gray!6}{0} & \cellcolor{gray!6}{1} & \cellcolor{gray!6}{0}\\
 & requests & 4 & 0 & 0 & 0 & 0 & 0 & 0 & 0 & 0 & 0 & 0 & 0 & 0 & 0 & 0 & 0 & 4 & 0\\
& \cellcolor{gray!6}{seaborn} & \cellcolor{gray!6}{0} & \cellcolor{gray!6}{0} & \cellcolor{gray!6}{\textbf{1}} & \cellcolor{gray!6}{0} & \cellcolor{gray!6}{0} & \cellcolor{gray!6}{\textbf{1}} & \cellcolor{gray!6}{0} & \cellcolor{gray!6}{\textbf{1}} & \cellcolor{gray!6}{0} & \cellcolor{gray!6}{\textbf{1}} & \cellcolor{gray!6}{0} & \cellcolor{gray!6}{\textbf{1}} & \cellcolor{gray!6}{\textbf{1}} & \cellcolor{gray!6}{0} & \cellcolor{gray!6}{\textbf{1}} & \cellcolor{gray!6}{\textbf{1}} & \cellcolor{gray!6}{2} & \cellcolor{gray!6}{1}\\
 & setuptools & 8 & 0 & 52 & 10 & 0 & 52 & 0 & 52 & 13 & 60 & 11 & 52 & 56 & 14 & \textbf{62} & 57 & 177 & 89\\
& \cellcolor{gray!6}{sunpy} & \cellcolor{gray!6}{0} & \cellcolor{gray!6}{0} & \cellcolor{gray!6}{\textbf{1}} & \cellcolor{gray!6}{0} & \cellcolor{gray!6}{0} & \cellcolor{gray!6}{\textbf{1}} & \cellcolor{gray!6}{0} & \cellcolor{gray!6}{\textbf{1}} & \cellcolor{gray!6}{0} & \cellcolor{gray!6}{\textbf{1}} & \cellcolor{gray!6}{0} & \cellcolor{gray!6}{\textbf{1}} & \cellcolor{gray!6}{\textbf{1}} & \cellcolor{gray!6}{0} & \cellcolor{gray!6}{\textbf{1}} & \cellcolor{gray!6}{\textbf{1}} & \cellcolor{gray!6}{11} & \cellcolor{gray!6}{1}\\
 & xonsh & 0 & \textbf{1} & 0 & 0 & 0 & \textbf{1} & \textbf{1} & 0 & \textbf{1} & \textbf{1} & \textbf{1} & 0 & 0 & 0 & \textbf{1} & 0 & 10 & 2\\
  \cmidrule{2-20}
& \cellcolor{gray!6}{12 Projects Total} & \cellcolor{gray!6}{12} & \cellcolor{gray!6}{2} & \cellcolor{gray!6}{54} & \cellcolor{gray!6}{12} & \cellcolor{gray!6}{0} & \cellcolor{gray!6}{56} & \cellcolor{gray!6}{1} & \cellcolor{gray!6}{54} & \cellcolor{gray!6}{20} & \cellcolor{gray!6}{65} & \cellcolor{gray!6}{14} & \cellcolor{gray!6}{55} & \cellcolor{gray!6}{64} & \cellcolor{gray!6}{20} & \cellcolor{gray!6}{\textbf{67}} & \cellcolor{gray!6}{65} & \cellcolor{gray!6}{260} & \cellcolor{gray!6}{99}\\
\midrule
\multicolumn{2}{l}{52 Projects Total} & 86 & 109 & 64 & 12 & 1 & 134 & 127 & 69 & 143 & 83 & 24 & 147 & 137 & 139 & 85 & \textbf{156} & 608 & 283\\
\bottomrule
\end{tabular}
\end{table*}

\section{Results}
\label{sec:results}

\subsection{Answering RQ1: \rqone}
\label{sec:results:rq1}

This research question evaluates prevalence of Resource-Affected Flaky Tests (RAFTs) among flaky tests. 
Table~\ref{lbl:rq1} summarizes the results of \numTestRuns{}~test runs on each of the \numProj{} projects for every throttling configuration in Table~\ref{tab:throttling-configs}.
For every project in the table, the column ``Baseline Flaky'' contains the number of flaky tests identified under no resource throttling. 
The columns ``Flaky'' and ``RAFTs'' under ``Total, All Runs'' represent the total number of flaky tests identified across all configurations, and those unique tests which can be considered RAFTs, respectively.
The remaining columns show how many RAFTs were observed by throttling different kinds of resources: CPU \cpuOnly{}, Memory \memOnly{}, Disk \diskOnly{}, or Network \networkOnly{}, and combinations thereof.

\subsubsection{\textbf{RQ1.1} \rqoneone{}} 
With no resource throttling, we observed a total of \numBaselineFlaky{}~flaky-test failures when running the tests of each project for \numTestRuns{}~times and aggregating results across all \numProj{} projects.
Across all configurations, we observed a total of \numTotalFlaky{}~flaky test failures, of which \numTotalRAFT{}~tests were classified as RAFTs.
The highest number of RAFTs identified in a single Java project is \numJavaWebsocketRAFT{}~in \CodeIn{java-websocket} (=\percentJavaWebsocketRAFT{}\% of the total of flaky tests on that project).
Within the JavaScript projects \CodeIn{apollo-client-devtools} showed the highest number of RAFTs (13, or 61.90\% of total flaky tests in that project), and within the Python projects, \CodeIn{setuptools} showed the highest number of RAFTs (89, or 50.28\% of the total flaky tests in that project). 
We observed no RAFTs in \numProjWithoutRAFT{} projects of \numProj{}~projects.
Note that the number of test runs for all combinations of resource throttling is significantly greater than that for no resource throttling (4,500=\numTestRuns{}*15 versus \numTestRuns{}). We run baseline configuration and every other configuration for \numTestRuns{}~times. 
This, in conjunction with potentially increased failures in RAFTs due to throttling accounts for the difference in the numbers between the columns ``Baseline Flaky'' and ``Total, All Runs/Flaky''. 


\begin{center}
\vspace{-1ex}
\begin{tcolorbox}[enhanced,width=0.45\textwidth,center upper,drop shadow southwest,sharp corners]
\emph{Summary:}~Of all flaky tests detected in our study, we find that \percentRAFT{}\%(=\numTotalRAFT{}/\numTotalFlaky{}) of them are RAFTs. 
\end{tcolorbox}
\vspace{-1ex}
\end{center}

\begin{figure}
\begin{subfigure}{0.55\textwidth}
\includegraphics[width=3.4in,trim=0 120 0 120,clip]{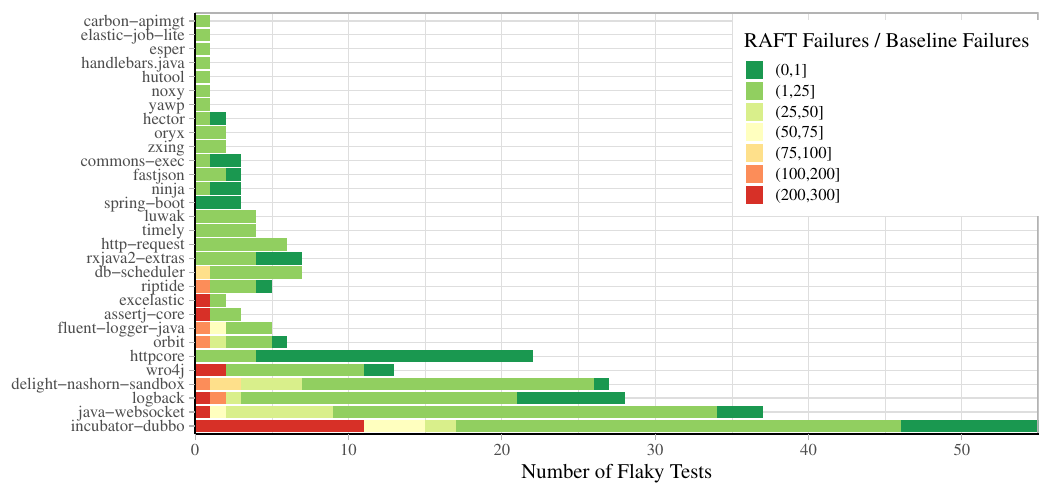}
\caption{Java Projects}
\end{subfigure}
~
\begin{subfigure}{0.55\textwidth}
\includegraphics[width=3.4in,trim=0 170 0 140,clip]{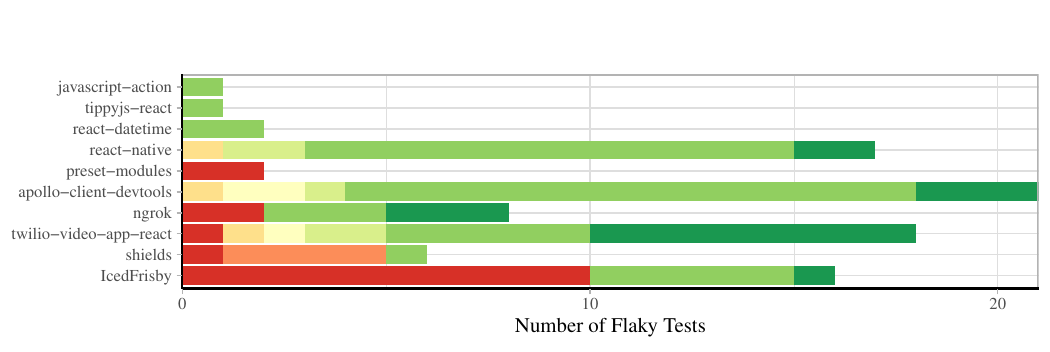}
\caption{JavaScript Projects}
\end{subfigure}
~
\begin{subfigure}{0.55\textwidth}
\includegraphics[width=3.4in,trim=0 170 0 140,clip]{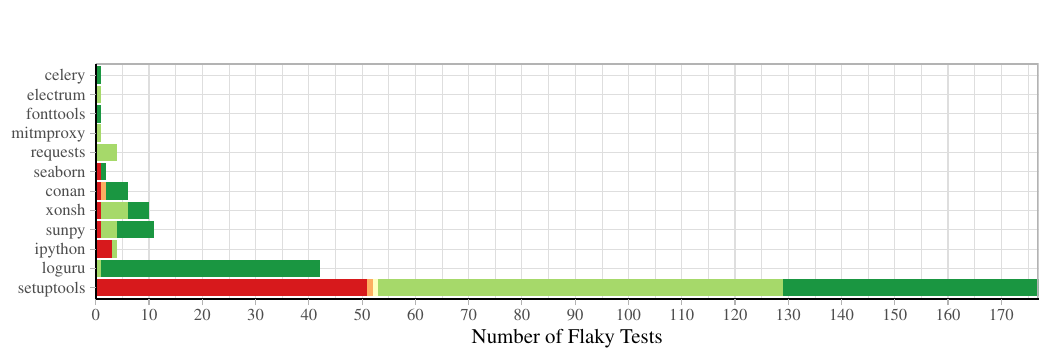}
\caption{Python Projects}
\end{subfigure}
\caption{\label{fig:rq12}Just how resource affected are these flaky tests? \textnormal{For each project with flaky tests, we show the failure increase rate from the baseline ``no throttling'' configuration to the most failure-inducing resource throttling condition.}}
\vspace{-1ex}
\end{figure}


\subsubsection{\textbf{RQ1.2} \rqonetwo{}} 
The barplot in Figure~\ref{fig:rq12} shows the distribution of tests on various ``resource-affectedness'' levels for each project that contains RAFTs.  A level is defined as the increase in failure rate relative to the baseline ``no throttling'' configuration. The colors indicate different levels. For example, dark green denotes a test that is not resource affected, while red indicates a test that is severely resource affected.
The length of each bar denotes the number of flaky tests found with throttling runs, i.e., it corresponds to the value in column ``Total, All Runs/Flaky'' on Table~\ref{lbl:rq1}. 

\sloppy
Based on Figure~\ref{fig:rq12}, we observe that the level of resource-affectedness varies with each project.
Most projects contain RAFTs which are slightly affected by resources with an increase in failure rate of less than 25x (shown in light green).
Some projects such as \CodeIn{incubator-dubbo} for Java, \CodeIn{IcedFrisby} for JavaScript, and \CodeIn{setupTools} for Python
have a large  number of RAFTs that are heavily affected by resource availability with an increase in failure rate of over 200x.
RAFTs that are heavily affected by resources are quite uncommon and mostly exist in small numbers in few projects.
We should note that RAFTs that are heavily affected by resources may already be well-known to developers, as they are clearly very sensitive to resource availability, and are quite likely to fail if resources are unavailable.
On the contrary, the most commonly-occurring RAFTs in our experiment (shown in light green, those that increased in failure rates more marginally), may be the most dangerous, since developers are less likely to make the connection between the flaky-test failures and resource availability.
Heavily affected RAFTs on the other hand, can fail two or hundreds of times more frequently than normal and are therefore easier to identify.
Section~\ref{sec:pull-requests} describes developer feedback to the RAFTs that we identify.


\begin{center}
\vspace{-1ex}
\begin{tcolorbox}[enhanced,width=0.45\textwidth,center upper,drop shadow southwest,sharp corners]
\emph{Summary:}~Most commonly, RAFTs are slightly affected by resources. In a 0-300 scale of resource-affectedness, the most predominant range is 1-50. 
\end{tcolorbox}
\vspace{-1ex}
\end{center}

\subsection{Answering RQ2: \rqtwo}

This research question evaluates the impact of individual resources and their combination on test flakiness.
Table~\ref{lbl:rq1} summarizes the test failures for \numTestRuns{}~runs on all configurations.
The columns \cpuOnly{}, \memOnly{}, \diskOnly{}, and \networkOnly{} contains the number of test failures for throttling of individual resources.
The columns after those show the test failures for combinations of these resources.

\subsubsection{\textbf{RQ2.1} \rqtwoone{}} 
We begin by examining the characteristics by language.
Across all \numProjJava{} Java projects, we observed \Num{82} test failures under CPU throttling, \Num{10} under memory throttling, \Num{one} under disk throttling, and \Num{one} under network throttling.
Across all \numProjJS{} JavaScript projects, we see a similar pattern: we observed \Num{28} failures under CPU throttling, \Num{zero} under memory throttling, \Num{1} under disk throttling, and \Num{1} under network throttling.
This trend breaks among the  \numProjPython{} Python projects, apparently due to the influence of a single project (setuptools): we observe only \Num{two} failures under CPU throttling, \Num{54} under memory throttling, \Num{13} under disk throttling, and \Num{zero} under network throttling.
This behavior might be explained by different minimum memory requirements across projects: our ``memory'' throttling configuration allows only 512MB of RAM, which was insufficient to even run tests for some of the projects with the most RAFTs in other languages (e.g. incubator-dubbo, apollo-client-dev-tools), whereas this amount appeared to be just enough to run the tests in setuptools, albeit sufficiently little to cause many flaky test failures.
However, what is clear from the results is that disk and network throttling play a much more minor role in causing flaky test failures.
Hence, we conclude that CPU starvation is the most significant and ubiquitous factor for increasing test failures, while memory throttling may also be impactful.


\begin{center}
\begin{tcolorbox}[enhanced,width=0.45\textwidth,center upper,drop shadow southwest,sharp corners]
\emph{Summary:}~The resource that triggers flakiness most frequently in Java and JavaScript projects is the CPU, and in Python is the memory.
\end{tcolorbox}
\end{center}

\subsubsection{\textbf{RQ2.2} \rqtwotwo{}} 
Recall that, according to our definition, a RAFT is a test that has a statistically greater failure rate in at least one resource-throttled configuration, as compared to its baseline failure rate (under no throttling).
Of the total of \numTotalRAFT{}, we identified \numTotalRAFTInOneConfig{} tests that were RAFTs in only one configuration.
In each case, the absolute difference in failures was relatively small, ranging from an increase between \Num{seven} and \Num{23} additional failures observed under the single configuration that exposed the test as RAFT and the baseline failure count.
To further investigate these tests, we use Pearson's chi-squared test to determine whether there were other throttling configurations that resulted in failure rates that were statistically indistinguishable from that single RAFT case.
In all but \Num{three} cases, we found at least one other throttling configuration that induced the RAFT to fail at a rate that was indistinguishable from both the RAFT-inducing configuration and the baseline configuration.
\Num{Two} of these tests belonged to the Python project \texttt{setuptools}, and were RAFTs only in the CMD configuration.
The last test belonged to the Java project \texttt{riptide}, and was flaky only in memory-throttling configurations, failing persistently in CPU-throttled configurations.
We conclude that it is unlikely that it is necessary to examine every resource configuration in order to detect RAFTs, and simply increasing the number of trials may be sufficient to increase confidence levels.
Nonetheless, in projects where tests are known to rely on disk input/output (such as in the case of the \texttt{setuptools} project), adding disk throttling combinations may help to detect RAFTs.

\begin{center}
\begin{tcolorbox}[enhanced,width=0.45\textwidth,center upper,drop shadow southwest,sharp corners]
\emph{Summary:}~RAFTs rarely manifest only in specific throttling configurations. Of the \numTotalRAFT{} RAFTs, \numTotalRAFTInOneConfig{} of them manifested only in one of the \Num{15} configurations. 
\end{tcolorbox}
\end{center}

\subsection{Answering RQ3: \rqthree}


\begin{figure*}
\includegraphics[width=7in,trim=0 165 0 165,clip]{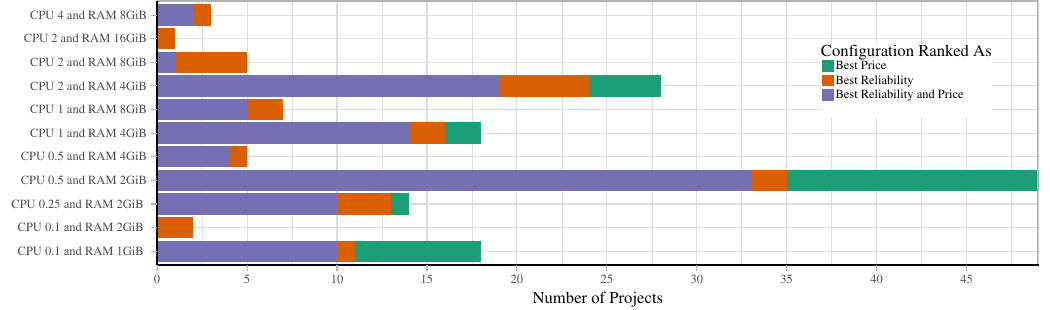}
\caption{\label{fig:rq3}What are the best resource configurations \underline{\smash{to prevent flaky failures}}? \textnormal{For each configuration that we analyzed, we show the number of times that it was the best at avoiding flaky failures, the best in terms of price, or the best in terms of both. If a configuration was tied for best in terms of reliability for a project, we select the cheaper one. We hide configurations that were not optimal on either dimension.}}
\vspace{-2ex}
\end{figure*}

This research question evaluates the reliability of test execution and cost for each AWS configuration.
To answer this question, we analyzed the percentage of build failures and the price to run every project on each AWS configuration in Table~\ref{tab:aws-configs}. 
Figure~\ref{fig:rq3} summarizes the results as a stacked bar chart ranking each configuration on price, reliability, and both.
Some projects may run with equal reliability on more than one configuration, but have different price points.
In such cases, the project is only shown under the ``Best Reliability and Price'' category.
Conversely, some projects may show significantly different behavior at drastically different price points on different configurations.

Figure~\ref{fig:rq3} shows that the best configuration for reliability and price largely depends on individual projects. We observe that the price of builds does not scale linearly with price of configurations.
Configurations with lower resources generally have a lower billing rate.
However, the constrained resources can make individual builds significantly slower.
As a result, it can often be more expensive to run builds on cheaper configurations compared to those with slightly higher billing rates.
In addition, some projects could not be run on low resources configurations due to catastrophic failures (e.g., deterministically running out of memory).
The most reliable and most cost effective configurations vary based on the projects.
We observed that the configuration ``CPU 0.5 and RAM 2GiB'' is the most cost effective configuration, followed by ``CPU 2 and RAM 4GiB'' and ``CPU 1 and RAM 4GiB.''
We observed that the configurations ``CPU 0.5 and RAM 2GiB'' and ``CPU 2 and RAM 4GiB'' are the most reliable configurations.
A table showing the price and reliability of each configuration for each project is included in the appendix to this article.

\OurComment{
There are (1) nondeterministic RAFTs (tests that fail statistically significantly more in different resource configurations), (2) deterministic RAFTs (tests that always fail 0 or 100\% depending on the configuration), (3) non-RAFT but flaky (tests that do NOT fail statistically significantly more in different resource configurations), and (4) non-flaky tests (tests that always passed or always failed in our experiments).
We seem to treat (2) and (4) as the same now.
}

\begin{center}
\begin{tcolorbox}[enhanced,width=0.45\textwidth,center upper,drop shadow southwest,sharp corners]
\emph{Summary:}~The most cost-effective configuration to prevent RAFTs largely depends on the project.
\end{tcolorbox}
\end{center}

\subsection{Answering RQ4: \rqfour}

\begin{figure*}
\includegraphics[width=7in,trim=0 185 0 175,clip]{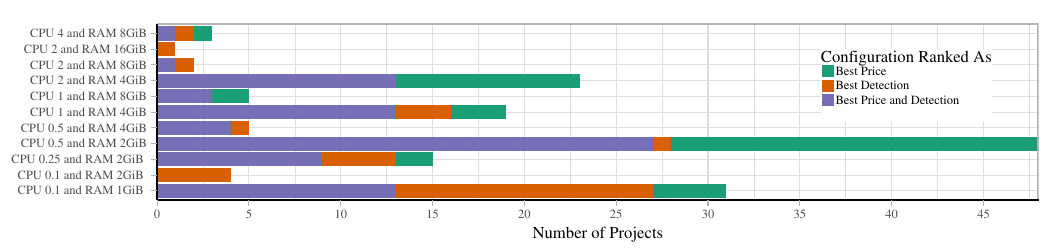}
\caption{\label{fig:rq4}What are the best resource configurations \underline{\smash{to detect flaky failures}}? \textnormal{For each configuration that we analyzed, we show the number of times that it was best at detecting flaky tests (number of unique flaky tests detected), the best in terms of price, or the best in terms of both. If a configuration was tied for best in terms of detection for a project, we select the cheaper one. We hide configurations that were not optimal on either dimension.}}
\vspace{-2ex}
\end{figure*}

This research question evaluates the reliability of test failures and cost for each AWS configuration.
To answer this question, we analyzed the number of test failures and the cost to run every project on each AWS configuration in Table~\ref{tab:aws-configs}.
Figure~\ref{fig:rq4} summaries the results as a stacked bar chart ranking each configuration on price, detection, and both.
As discussed in the prior sub-section, the cost for individual runs depends on the time taken for each run and thus can vary significantly for different configurations.
The cost of execution on cheaper configurations can be more than the cost of execution on expensive configurations because of the longer execution time on weaker hardware.
Similar to RQ3, Figure~\ref{fig:rq4} shows that the best configuration for detecting flaky test failures and obtaining best price depends on individual projects. Note that the ``best price'' configuration is the same for both detecting or avoiding flakiness.

The most cost effective and failure detecting configurations vary based on the projects.
We observe that the configuration ``CPU 2 and RAM 4'' is cost effective on many projects, but not every effective at detecting flaky failures.
We observe that the configuration ``CPU 0.1 and RAM 1GiB'' is effective at detecting test failures on many projects but not very cost effective.
However, the configuration ``CPU 0.5 and RAM 2GiB'' combines the desirable properties of the previously discussed configurations for our dataset and is the most cost effective and reliable configuration for detecting flaky tests.

\begin{center}
\vspace{-1ex}
\begin{tcolorbox}[enhanced,width=0.45\textwidth,center upper,drop shadow southwest,sharp corners]
\emph{Summary:}~The most cost-effective configuration to detect RAFTs largely depends on the project.
\end{tcolorbox}
\vspace{-1ex}
\end{center}

\section{Discussion}
\label{sec:discussion}
\label{sec:implications-developers}
Our study confirms the presence of resource-affected flaky tests (RAFTs) in open-source Java, JavaScript, and Python projects.
Our experiments\Space{, perhaps not surprisingly,} find that the presence and impact of RAFTs can vary substantially between projects.
While some of these failures might be obvious --- in the case of \CodeIn{incubator-dubbo}, we find tests that failed in almost every single run when executed with restricted resources --- other tests have a more subtle dependency on system resources.
As test suites grow, and resources are increasingly stretched thin to run more test suites concurrently, developers should be aware of slowly-increasing flaky-test failure rates.

Projects that have mostly small, deterministic tests are less likely to be impacted by resource-related flakiness than other projects that have large, resource-dependent integration tests.
By specifying the expected resource requirements for reliably running tests, developers can reduce the occurrence of RAFT-related failures.
In this section, we describe our experiences reporting these concerns to developers, provide a qualitative discussion of exemplar and unusual RAFT and non-RAFT flaky tests, and implications for future research in flaky tests.

\subsection{Feedback from developers}
\label{sec:pull-requests}


To gain further insight into the implications of our study, we contacted developers of projects in which we detected RAFTs and suggested they update the project specifications (e.g., \CodeIn{README.md}) with the minimal resource configurations that should be used to mitigate RAFTs.
As developers are likely more hesitant to accept changes related to their projects' specifications (than typical, code-related changes), we open issue reports on only a subset of our evaluation projects to gauge developers' interest in our findings -- specifically we start by engaging with developers only for our Java-based projects.
We next describe the process by which we contacted developers, along with a summary of the developers' responses to our recommendations.
\OurComment{Removed by Wing as we now explain why we only use a subset of projects. Denini will address the remaining ones while this work is under review.
\jon{Denini: Did you ever follow-up on the Python/JS projects? What should we say here to explain why we only have contacted developers of Java projects?}
\Den{@Jon: I don't remember exactly, but it seems that we didn't make contact with the python/js developers because we had few or zero RAFTs in the projects. (I tried looking in the slack history, but it disappeared).}
\Den{Maybe if there are few, and we have time I can do it...}
}

\subsubsection{Overview}
First, we identify Java projects from our evaluation that:
\begin{itemize}
    \item \textit{are active}, i.e., had a commit or developer interaction on issues or pull requests in the last three years (this resulted in only \Num{18} of the \numProjJava{} projects);
    \item \textit{are runnable}, i.e., could be cloned at their latest version, the code compiles, and the tests run (all \Num{18} ran);
    \item \textit{had a flaky test in their latest revision}, i.e., we run the project test suites \numTestRuns{} times in each of the \Num{12} throttling configurations from Table~\ref{tab:aws-configs}, and keep the projects that have at least one RAFT
    (\numProjRAFTContact{} of the \Num{18} projects had at least one RAFT). The latest version of each project that we used is in our artifact~\cite{raft-dataset}.\OurComment{Table~\ref{tab:results-latest} summarizes this result.} 
    \OurComment{Wing: Agreed. Denini should run the check to confirm which of these flaky tests are RAFTs. In the meantime, we should remove the table.
    \jon{Denini: Is the column there the number of flaky tests, or the number of RAFT? Shouldn't we be only looking at RAFT?}
    \Den{@Jon, I made pr/issues showing developers all the flaky we found, that's why I made this table showing the number of flaky tests. As you mentioned in the table, perhaps it would be better to delete it.}
    }
\end{itemize}

For each of the remaining \numProjRAFTContact{} projects, we initiate communication with developers by creating an issue, either via GitHub or on the project's custom issue tracking system, that clearly indicates which tests are RAFTs, how to reproduce the issue (with step-by-step instructions on how to clone the code, build the Docker container, and run the tests), and which configuration likely prevents RAFTs.

\OurComment{
\begin{table}[t!]
\jon{What is the purpose of this table? Does it matter how many flaky tests we found in this experiment? If it matters, we shoudl talk about the results. Otherwise, I advocate for removing the table.}
\caption{Results of running the projects in latest version.}
\label{tab:results-latest}
\centering
\begin{tabular}{lrr}
\toprule
\textbf{project}        &  \textbf{sha} &  \textbf{\# flaky test} \\
\midrule
assertj                 & 9b653c5 &  5\\
commons-exec            & 7204a35  &  0\\
db-scheduler            & cc67408 &  3\\
delight-nashorn-sandbox & ac868e1 &  3\\
dubbo                   & acd4212 &  7\\
elasticjob              & 2bfce1e &  13\\
esper                   & 8d172d3 &  0\\
fastjson                & c942c83 &  1\\
handlebars.java         & 683c5e8 &  0\\
http-core               & 7376337 &  4\\
hutool                  & 89832c &  0\\
java-websocket          & 30ba037 &  20\\
logback                 & 6abce2f &  27\\
ninja                   & fa8c86 &  1\\
riptide                 & 1986672 &  13\\
rxjava2-extras          & 3f35966 &  6\\
timely                  & 1443fdc &  4\\
zxing                   & bb75858 &  1\\
\midrule
\textbf{Total}          & \textbf{-} &  \textbf{108}\\
\end{tabular}
\end{table}
}

\subsubsection{Feedback}

For \numProjRAFTContact{} projects that we initiated communication with, \numProjRAFTResponded{} have responded to us and the remaining \numProjRAFTPending{} are still pending with no response.
In \Num{two} of the \numProjRAFTResponded{} projects~\cite{dubbo-12391, delight-nashorn-sandbox-137}, developers quickly asked for or created their own PRs with minimum required resources to avoid RAFTs after we reached out.
In \Num{two} other projects~\cite{http-core-747-issue, db-scheduler-388-issue}, developers initially expressed confusion to the concept of RAFTs and did not understand why stating minimum requirements for running tests were relevant.
However, after follow-up comments describing how the RAFT configurations were plausible cases in which a contributor might run their tests, the developers agreed with our concern and accepted our proposed documentation changes.
Lastly, there were \numProjRAFTRespondedFix{} projects~\cite{timely-242, shardingsphere-elasticjob-2219} in which developers felt that the change to documentation was not appropriate, noting that these tests might be flaky ``regardless of resources'' and preferred to create a task to improve those tests to reduce their flakiness directly.



\subsubsection{Discussion}

There were common threads throughout our interactions with developers.

First, developers generally want to reduce the flakiness in their test suites, but are more focused on fixing tests to make them less flaky.
In {\tt http-core}~\cite{http-core-747-issue, http-core-410}, the developer was initially quite dismissive, and eventually acquiesced with the caveat that they would ``rather see efforts spent on analyzing the failing test cases and fixing them''.
The developer in {\tt shardingsphere-elasticjob}~\cite{shardingsphere-elasticjob-2219} simply linked our issue to another flaky-test related issue, and eventually fixed the flakiness by using more robust assertions.
These developers were not convinced that poorly provisioned machines can lead to more flakiness, perhaps because they do not experience the flakiness on their own machines or believe that their application would be used on such machines.

In fact, developers of the {\tt timely}~\cite{timely-242} project had difficulties imagining their applications/test suites would be run in resource-constrained environments.
The developer was not interested in specifying minimum resources because they said their application is ``designed to work with Apache Accumulo, an inherently `big data' application''.
Similarly, in {\tt db-scheduler}~\cite{db-scheduler-403, db-scheduler-388-issue}, the developer said that they always ``run the tests on a multicore machine''.
In these situations, noting that different developers have different machine specifications helped.
(Note that suggesting asynchronously executing test suites in cloud environments may also have convinced developers, though we did not try this suggestion.)

There are also developers like those in {\tt delight-nashorn-sandbox}~\cite{delight-nashorn-sandbox-137} and {\tt dubbo}~\cite{dubbo-12391} that were immediately interested in our fixes.
In fact, the developer of the former created a new test that failed if system specifications were not adequate to better inform developers that RAFT failures may be occurring. This example highlights how developers may be content with simple RAFT mitigation strategies.



In summary, we reached out to the developers of \numProjRAFTContact{} Java projects regarding RAFTs and the minimum machine specifications their projects' should run in to avoid RAFTs.
Of the \numProjRAFTContact{} projects we reached out to, developers of \numProjRAFTResponded{} projects have responded to us, while the remaining \numProjRAFTPending{} are pending with no response.
Of the \numProjRAFTResponded{} that responded, developers of \numProjRAFTRespondedSpecs{} projects have improved their project based on our suggestions, while the remaining \numProjRAFTRespondedFix{} indicated that they prefer actual fixes over specification clarifications.

\subsection{Qualitative examination of flaky tests}
To complement our statistical analysis, we also provide a qualitative discussion of exemplar and unusual RAFT and non-RAFT flaky tests.
\subsubsection{An unusual RAFT case}


Intuitively, a RAFT is due to the limited availability of resources (e.g., lack of RAM needed to properly run the test).
However, we also observed RAFTs that are due to the abundance of resources.
One such example is the test \CodeIn{Issue677}~\cite{java-websockets-test677} from the project \CodeIn{Java-Websocket}.
This test first creates a server and two clients, instances of the \CodeIn{WebSocketServer} and \CodeIn{WebSocketClient} classes, respectively. Then, the test starts the server and connects the first client to the server. After connecting, the test calls \CodeIn{webSocket0.close()} under the thread of the first client to close the connection to the server.
The test then checks if the server is indeed closing. 
However, sometimes, when the assertion runs, the connection with the server has already been closed (when the machine is highly available) or it is still open (when the machine is highly overloaded), as opposed to ``being closed.''
As a result, the assertion in this test fails when the machine is overloaded or when it is highly available.
In more detail, this test failed \Num{14} times out of \numTestRuns{} executions in a regular execution where resources have not been throttled (4CPU and 16GiB of RAM). 
However, when we run this test in an environment with slightly less CPU (between 0.5 and 4), memory (between 4GiB and 16GiB), and disk I/O throttling, the test can actually fail less often than the regular execution -- failing as little as just \Num{four} times out of \numTestRuns{} executions.
On the other hand, when the test is run with less than 0.1CPU or 2GiB of RAM, the test always failed more \Num{30} times and can even fail up to \Num{69} times. 
We submitted a pull request to the owner of the project explaining why that specific assertion was unreliable and recommended its removal (the test has other assertions). The owner accepted the PR.

\subsubsection{Non-RAFT flaky tests}
Of course, not all flaky tests are RAFTs.
We provide a qualitative discussion of two flaky tests that we observed in our dataset, which are not RAFTs.
The {\tt Timely}~\cite{GitHubTimely} project contains two tests 
from the class \CodeIn{Time\-Series\-Grouping\-Iterator\-Test} that are flaky but non-RAFT: \CodeIn{test\-Time\-Series\-Drop\-Off} and \CodeIn{test\-Multiple\-Time\-Series\-Moving\-Average}. 
The goal of these tests is to check that the averages of numeric values in two data structures are the same. 
However, the tests use time-based random number generation, which results in unpredictable and unreliable test results.


The flakiness of these tests is not dependent on the environment in which they are run, but rather on the time in which they are run. 
A developer can run the same test multiple times, even in different environments, and get different results each time. 
In our experiments, we observed that these tests were flaky in all configurations. 
In the baseline configuration, the tests failed \Num{eight} times in \numTestRuns{} runs, while in other configurations, the tests failed at least once and up to \Num{12} times. 
Our findings for this test are confirmed by our prior anlaysis~\cite{Lam2020WASM}, which also described these tests to be flaky due to time.
In fact, that prior analysis found a \Num{2.6\%} failure rate for these tests, which translates to roughly \Num{eight} failures in \numTestRuns{} runs.

\subsection{Implications for researchers}
\label{sec:implications-researchers}
One line of flaky test research has focused on \emph{detecting} flaky tests, generally by re-running them hundreds~\cite{Lam2019OSMX} or thousands~\cite{Alshammari2021MHB,Lam2020WASM} of times to detect unlikely failures.
We have found that these experiments can be conducted more cost effectively by reducing the resources that are used for each test execution: providing each test suite with 4 CPUs and 16GiB of RAM may be an over-provisioning of resources.
From our experiments, we found that as long as there is enough RAM available to reliably complete a test suite execution without reaching a fatal out-of-memory error, reducing resources available to a test suite increases the number of flaky tests detected.
Rather than deploying ``stressor'' tasks that acquire CPU and RAM in an effort to starve tests of resources~\cite{Silva2020TA,Terragni2020SF}, it may be substantially cheaper to simply limit the resources available to those tests, and use those resources for other\Space{ productive} purposes. 

Surveys of developers show that detection of flaky tests is a less pressing problem than the mitigation of flaky tests~\cite{Gruber2022F}.
This article outlines a simple, yet effective approach for mitigating the impact of flakiness in test suites, when that flakiness is tied to resource availability.
Researchers should investigate other approaches to reduce the incidence of flaky failures, considering factors beyond the test code itself, such as environmental factors.
Our supplemental artifact includes our dataset and the scripts used to detect RAFTs from test executions~\cite{raft-dataset}.


\subsection{Implications for Continuous Integration Infrastructure}
Cloud-based continuous integration (CI) systems have become increasingly popular.
Major cloud vendors provide CI services, such as Amazon's CodeBuild~\cite{awsCodeBuild}, Microsoft's Azure DevOps Pipelines~\cite{azureDevOps} and Google Cloud Build~\cite{googleCloudBuild}.
We reviewed the pricing and configuration options available for popular cloud-based CI services, to see how the configurations aligned with the resource configurations that we evaluated.
Specifically, we reviewed the configuration and pricing of Amazon's CodeBuild~\cite{awsCodeBuild}, Microsoft's Azure DevOps Pipelines~\cite{azureDevOps}, Google Cloud Build~\cite{googleCloudBuild}, GitHub Actions~\cite{gitHubActionsRunners}, GitLab CI/CD~\cite{gitLabRunners}, BitBucket Pipelines~\cite{bitBucketRunners}, CircleCI~\cite{circleCIRunners}, TravisCI~\cite{travisBuild} and TeamCity~\cite{teamCity}.
Builds are executed by \emph{CI runners}, which may be provided by the cloud service (``cloud builders''), or managed by developers using their own (``self-hosted builders'').
Some services provided only a single configuration of cloud builder: Azure DevOps and GitHub Actions both provide runners with 2 CPUs and 7GiB of RAM~\cite{azureDevOps,gitHubActionsRunners} (at time of writing, GitHub has a beta-only feature to support larger cloud runners).
BitBucket provides an unspecified CPU resource, but allows memory to be scaled between 4 and 32 GiB~\cite{bitBucketRunners}.
GitLab and Google Cloud Build allow developers to select as little as 1 CPU with 4GiB of RAM~\cite{gitLabRunners,googleCloudBuild}, while Amazon CodeBuild, GitLab CI, TravisCI and TeamCity start at 2 CPUs with 4GiB of RAM, with a maximum configuration (on Amazon CodeBuild) of 72 CPUs and 144GiB of RAM~\cite{awsCodeBuild,gitLabRunners,travisBuild,teamCity}.

Our finding that some projects can reliably build with only 0.5 CPU and 2GiB of RAM indicates that some developers may be able to save money by using lower-end CI runners than are available using the ``cloud runner'' model.
Each of these CI services \emph{also} supports a ``self-hosted'' runner model, where builds take place on compute resources that are managed by the developers (e.g., a dedicated ``builder'' machine, or an auto-scaling cluster of builders).
For example: developers could deploy an auto-scaling cluster of builders with 0.5CPU/2GiB RAM on AWS for \$0.008739/hour, while GitHub Actions would charge about 55 times as much (\$0.008/\emph{minute}) for a runner with 2 cores and 7GiB of RAM.
Furthermore, each of these services support only a limited number of resource configurations, forcing specific combinations of CPU and memory (e.g., on CircleCI, a configuration 1 CPU with 4GiB of RAM is not available, developers must pay for 2 CPUs to receive 4GiB of RAM).
Based on the mismatch between the resource requirements of projects and the configurations provided by cloud CI services, we believe that significant cost savings may be achievable for developers.
Using ``self-hosted'' runners that auto-scale on containers~\cite{githubActionsAutoScale} that match the actual resources required by a build (rather than over-provisioning) can have significant cost savings.





\subsection{Threats to validity}
\label{sec:threats} 



\subsubsection{Construct validity}

There are two central constructs to our study, the flakiness of tests and resource-dependency.

\emph{Test flakiness:} For classifying a test as flaky, we rely on the observation of different test outcomes across repeated executions.
These executions may be affected by uncontrolled factors and, hence, we may erroneously classify tests as flaky.
More precisely, there may exist an execution environment under which the tests do not non-deterministically pass and fail.
However, the manifestation of this behavior demonstrates that the investigated tests \emph{can} be flaky in \emph{some} execution environment.
This interpretation of flakiness follows common practice in existing work.

\emph{Resource dependency:} In our work, we determine resource dependency by controlling resource access via Linux control groups in a uniform manner, i.e., resource accesses are affected throughout test execution.
This access control closely resembles execution on a resource-constrained machine.
It does not resemble resource constraints resulting from dynamic load on shared resources well, as they are proposed in Terragni et al.'s work \cite{Terragni2020SF}.
However, the extreme resource restrictions we experimented with for Phase I of our work resemble extreme dynamic loads and lead to very sensitive detection.
We consider that part of our experiments to be a desirable property for a detector of rare events like flaky test failures.
For Phase II, the uniform resource restriction leads to potentially optimistic results if control for additional load on the test setup cannot be controlled for, which is an important restriction that users of our results should be aware of.

\subsubsection{Internal validity}

The conclusions we draw are based on \numTestRuns{}~re-executions of tests under each configuration.
Other work has shown that flaky tests can fail much more infrequently than once in \numTestRuns{}~runs \cite{Alshammari2021MHB} and our results do not systematically address such rare cases.
The focus of our work lies on flaky tests that fail frequently enough to significantly disrupt developer activity.

To classify tests as RAFTs, we rely on a $\chi{}^2$ test of independence between failure rates under normal operation and resource constraints.
As we consider different configurations for analyzing the effect of resource constraints, we conduct several such tests against the same baseline failure rate, which may lead to multiple comparison problems.
We account for these by adjusting the obtained p-values using the Benjamini-Hochberg procedure \cite{BenjaminiHochberg}.

\subsubsection{External validity}

Our results are restricted to the studied projects and may not generalize to other projects.
We expect that the main conclusions of this paper (that RAFTs exist, and that failure rates of RAFTs can be influenced by adjusting resource constraints) will hold.
However, it would be difficult to extrapolate from our study to determine precisely how prevalent RAFTs are in software overall, or which resource configurations are the ``best'' overall for reducing or increasing those failure rates.
Even from our study of only \numProj{} open-source projects, we can see that the observed prevalence of flaky tests and their sensitivity to resource restrictions differs.
We, hence, recommend to reassess RAFTs for other projects using the methodology outlined in this work.

\section{Conclusions}
\label{sec:conclusions}
Using rigorous statistical methods, we have empirically demonstrated the link between test flakiness and the resources available for running tests.
Our study of \numProj{} Java, JavaScript, and Python open-source projects revealed that resource-affected flaky tests (RAFTs) may be more prevalent in some projects than others, likely tied to the kinds of behaviors that each projects' test suite examines.
By controlling the quantity of CPU cores and RAM available to a test suite while it runs, developers can reduce the likelihood of observing flaky failures, or if desired, increase it.
When we reached out to the developers of \numProjRAFTContact{} projects regarding the minimum machine specifications their projects' should run in to avoid RAFTs, developers of \numProjRAFTResponded{} projects responded, while the remaining \numProjRAFTPending{} are pending with no response.
Of the \numProjRAFTResponded{} that responded, developers of \numProjRAFTRespondedSpecs{} projects improved their project based on our suggestions, while the remaining \numProjRAFTRespondedFix{} indicated that they prefer actual fixes over specification clarifications.
Comparing the cost of each cloud configuration, we found that developers can likely save money \emph{and} reduce flakiness by using a ``self-hosted'' CI runner configuration, as opposed to the ``cloud runners'' supported out-of-the-box by platforms like GitHub Actions.
Future research in detecting flaky tests will benefit from running tests in reduced resource configurations, which may be cheaper to run and reveal more flaky failures.
Future research on RAFTs might consider examining (1)~the different failures that occur under different resource configurations for a given test, (2)~the impact of other environmental factors on flaky-test failures, (3)~the idea of ignoring test runs when there are insufficient resources to reliably run tests, and (4)~how regression testing techniques, such as test parallelization, can leverage RAFT information to allocate machines for testing. 

\section*{Acknowledgements}
\noindent This work was supported in part by the National Science Foundation under grants NSF CCF-2100037 and CNS-2100015. 
 \bibliographystyle{IEEEtran}
\bibliography{references,wing-bibs/flaky-tests-links,wing-bibs/flaky-tests-papers,wing-bibs/crossrefs}

\begin{thebibliography}{10}
\providecommand{\url}[1]{#1}
\csname url@samestyle\endcsname
\providecommand{\newblock}{\relax}
\providecommand{\bibinfo}[2]{#2}
\providecommand{\BIBentrySTDinterwordspacing}{\spaceskip=0pt\relax}
\providecommand{\BIBentryALTinterwordstretchfactor}{4}
\providecommand{\BIBentryALTinterwordspacing}{\spaceskip=\fontdimen2\font plus
\BIBentryALTinterwordstretchfactor\fontdimen3\font minus
  \fontdimen4\font\relax}
\providecommand{\BIBforeignlanguage}[2]{{%
\expandafter\ifx\csname l@#1\endcsname\relax
\typeout{** WARNING: IEEEtran.bst: No hyphenation pattern has been}%
\typeout{** loaded for the language `#1'. Using the pattern for}%
\typeout{** the default language instead.}%
\else
\language=\csname l@#1\endcsname
\fi
#2}}
\providecommand{\BIBdecl}{\relax}
\BIBdecl

\bibitem{Luo2014HEM}
Q.~Luo, F.~Hariri, L.~Eloussi, and D.~Marinov, ``An empirical analysis of flaky
  tests,'' in \emph{FSE}, 2014.

\bibitem{Eck2019PCB}
M.~Eck, F.~Palomba, M.~Castelluccio, and A.~Bacchelli, ``Understanding flaky
  tests: {T}he developer's perspective,'' in \emph{ESEC/FSE}, 2019.

\bibitem{Parry2022KHMTOSEM}
O.~Parry, G.~M. Kapfhammer, M.~Hilton, and P.~McMinn, ``A survey of flaky
  tests,'' \emph{TOSEM}, 2021.

\bibitem{Kowalczyk2020NGSLM}
E.~Kowalczyk, K.~Nair, Z.~Gao, L.~Silberstein, T.~Long, and A.~Memon,
  ``Modeling and ranking flaky tests at {A}pple,'' in \emph{ICSE SEIP}, 2020.

\bibitem{Malm2020MJ}
J.~Malm, A.~Causevic, B.~Lisper, and S.~Eldh, ``Automated analysis of
  flakiness-mitigating delays,'' in \emph{AST}, 2020.

\bibitem{Rehman2021R}
M.~H.~U. Rehman and P.~C. Rigby, ``Quantifying no-fault-found test failures to
  prioritize inspection of flaky tests at {E}ricsson,'' in \emph{ESEC/FSE
  Industry Track}, 2021.

\bibitem{ResearchFacebookTesting}
\BIBentryALTinterwordspacing
``Facebook testing and verification request for proposals,'' {A}ccessed 2023.
  [Online]. Available:
  \url{https://research.fb.com/programs/research-awards/proposals/facebook-testing-and-verification-request-for-proposals-2019}
\BIBentrySTDinterwordspacing

\bibitem{Harman2018H}
M.~Harman and P.~O'Hearn, ``From start-ups to scale-ups: Opportunities and open
  problems for static and dynamic program analysis,'' in \emph{SCAM}, 2018.

\bibitem{GoogleTestingToT}
\BIBentryALTinterwordspacing
Google, ``Tot{T}: Avoiding flakey tests,'' {A}ccessed 2023. [Online].
  Available:
  \url{http://googletesting.blogspot.com/2008/04/tott-avoiding-flakey-tests.html}
\BIBentrySTDinterwordspacing

\bibitem{Memon2017GNDNSM}
A.~Memon, Z.~Gao, B.~Nguyen, S.~Dhanda, E.~Nickell, R.~Siemborski, and
  J.~Micco, ``Taming {G}oogle-scale continuous testing,'' in \emph{ICSE SEIP},
  2017.

\bibitem{Micco2017}
J.~Micco, ``{The state of continuous integration testing at Google},'' in
  \emph{ICST}, 2017.

\bibitem{Ziftci2017R}
C.~Ziftci and J.~Reardon, ``Who broke the build?: {A}utomatically identifying
  changes that induce test failures in continuous integration at {G}oogle
  scale,'' in \emph{ICSE}, 2017.

\bibitem{Jiang2017LYX}
H.~Jiang, X.~Li, Z.~Yang, and J.~Xuan, ``What causes my test alarm? {A}utomatic
  cause analysis for test alarms in system and integration testing,'' in
  \emph{ICSE}, 2017.

\bibitem{Herzig2015GCM}
K.~Herzig, M.~Greiler, J.~Czerwonka, and B.~Murphy, ``The art of testing less
  without sacrificing quality,'' in \emph{ICSE}, 2015.

\bibitem{Herzig2015N}
K.~Herzig and N.~Nagappan, ``Empirically detecting false test alarms using
  association rules,'' in \emph{ICSE}, 2015.

\bibitem{Lam2019GNST}
W.~Lam, P.~Godefroid, S.~Nath, A.~Santhiar, and S.~Thummalapenta, ``Root
  causing flaky tests in a large-scale industrial setting,'' in \emph{ISSTA},
  2019.

\bibitem{Lam2020MST}
W.~Lam, K.~Mu{\c{s}}lu, H.~Sajnani, and S.~Thummalapenta, ``{A study on the
  lifecycle of flaky tests},'' in \emph{ICSE}, 2020.

\bibitem{Leesatapornwongsa22flakerepro}
T.~Leesatapornwongsa, X.~Ren, and S.~Nath, ``Flake{R}epro: {A}utomated and
  efficient reproduction of concurrency-related flaky tests,'' in
  \emph{ESEC/FSE}, 2022.

\bibitem{DeveloperTestVerification}
\BIBentryALTinterwordspacing
``{Test verification},'' {A}ccessed 2023. [Online]. Available:
  \url{https://developer.mozilla.org/en-US/docs/Mozilla/QA/Test_Verification}
\BIBentrySTDinterwordspacing

\bibitem{Rahman2018R}
M.~T. Rahman and P.~C. Rigby, ``The impact of failing, flaky, and high failure
  tests on the number of crash reports associated with {F}irefox builds,'' in
  \emph{ESEC/FSE}, 2018.

\bibitem{silva2020shake}
D.~Silva, L.~Teixeira, and M.~d’Amorim, ``Shake it! detecting flaky tests
  caused by concurrency with shaker,'' in \emph{2020 IEEE International
  Conference on Software Maintenance and Evolution (ICSME)}.\hskip 1em plus
  0.5em minus 0.4em\relax IEEE, 2020, pp. 301--311.

\bibitem{luo-etal-fse2014}
\BIBentryALTinterwordspacing
Q.~Luo, F.~Hariri, L.~Eloussi, and D.~Marinov, ``An empirical analysis of flaky
  tests,'' ser. FSE 2014, pp. 643--653. [Online]. Available:
  \url{http://doi.acm.org/10.1145/2635868.2635920}
\BIBentrySTDinterwordspacing

\bibitem{raft-dataset}
D.~Silva, M.~Gruber, S.~Gokhale, E.~Arteca, A.~Turcotte, M.~D'Amorim, W.~Lam,
  S.~Winter, and J.~Bell, ``{The Effects of Computational Resources on Flaky
  Tests (Artifact)},'' \url{https://zenodo.org/doi/10.5281/zenodo.10015434},
  Oct. 2023.

\bibitem{Lam2019OSMX}
W.~Lam, R.~Oei, A.~Shi, D.~Marinov, and T.~Xie, ``{iDF}lakies: {A} framework
  for detecting and partially classifying flaky tests,'' in \emph{ICST}, 2019.

\bibitem{Shi2019LOXM}
A.~Shi, W.~Lam, R.~Oei, T.~Xie, and D.~Marinov, ``i{F}ix{F}lakies: {A}
  framework for automatically fixing order-dependent flaky tests,'' in
  \emph{ESEC/FSE}, 2019.

\bibitem{Shi2016GLM}
A.~Shi, A.~Gyori, O.~Legunsen, and D.~Marinov, ``Detecting assumptions on
  deterministic implementations of non-deterministic specifications,'' in
  \emph{ICST}, 2016.

\bibitem{Gruber2021LKF}
M.~Gruber, S.~Lukasczyk, F.~Kroi{\ss}, and G.~Fraser, ``An empirical study of
  flaky tests in {P}ython,'' in \emph{ICST}, 2021.

\bibitem{Terragni2020SF}
V.~Terragni, P.~Salza, and F.~Ferrucci, ``A container-based infrastructure for
  fuzzy-driven root causing of flaky tests,'' in \emph{ICSE NIER}, 2020.

\bibitem{GitHubDelightNashorn}
\BIBentryALTinterwordspacing
``{Nashorn Sandbox},'' {A}ccessed 2023. [Online]. Available:
  \url{https://github.com/javadelight/delight-nashorn-sandbox}
\BIBentrySTDinterwordspacing

\bibitem{Alshammari2021MHB}
A.~Alshammari, C.~Morris, M.~Hilton, and J.~Bell, ``{FlakeFlagger}:
  {P}redicting flakiness without rerunning tests,'' in \emph{ICSE}, 2021.

\bibitem{LamETAL20ISSRE}
W.~Lam, S.~Winter, A.~Astorga, V.~Stodden, and D.~Marinov, ``Understanding
  reproducibility and characteristics of flaky tests through test reruns in
  {J}ava projects,'' in \emph{ISSRE 2020: 31st IEEE International Conference on
  Software Reliability Engineering}, Virtual Event, October 2020, pp. 403--413.

\bibitem{Parry22Evaluating}
O.~Parry, G.~M. Kapfhammer, M.~Hilton, and P.~McMinn, ``Evaluating features for
  machine learning detection of order- and non-order-dependent flaky tests,''
  in \emph{2022 IEEE Conference on Software Testing, Verification and
  Validation (ICST)}, 2022, pp. 93--104.

\bibitem{Parry23Empirically}
\BIBentryALTinterwordspacing
------, ``Empirically evaluating flaky test detection techniques combining test
  case rerunning and machine learning models,'' \emph{Empir. Softw. Eng.},
  vol.~28, no.~3, p.~72, 2023. [Online]. Available:
  \url{https://doi.org/10.1007/s10664-023-10307-w}
\BIBentrySTDinterwordspacing

\bibitem{barbosa2022test}
K.~Barbosa, R.~Ferreira, G.~Pinto, M.~d'Amorim, and B.~Miranda, ``{Test
  Flakiness Across Programming Languages},'' \emph{IEEE Transactions on
  Software Engineering}, vol.~49, no.~4, pp. 2039--2052, 2022.

\bibitem{yostFlakyTestJSThesis}
G.~A. Yost \emph{et~al.}, ``{Finding flaky tests in JavaScript applications
  using stress and test suite reordering},'' Ph.D. dissertation, 2023.

\bibitem{Arteca22NPMFilter}
\BIBentryALTinterwordspacing
E.~Arteca and A.~Turcotte, ``Npm-filter: Automating the mining of dynamic
  information from npm packages,'' in \emph{Proceedings of the 19th
  International Conference on Mining Software Repositories}, ser. MSR
  '22.\hskip 1em plus 0.5em minus 0.4em\relax New York, NY, USA: Association
  for Computing Machinery, 2022, p. 304–308. [Online]. Available:
  \url{https://doi.org/10.1145/3524842.3528501}
\BIBentrySTDinterwordspacing

\bibitem{Yoo03Slurm}
A.~B. Yoo, M.~A. Jette, and M.~Grondona, ``Slurm: Simple linux utility for
  resource management,'' in \emph{Job Scheduling Strategies for Parallel
  Processing}, D.~Feitelson, L.~Rudolph, and U.~Schwiegelshohn, Eds.\hskip 1em
  plus 0.5em minus 0.4em\relax Berlin, Heidelberg: Springer Berlin Heidelberg,
  2003, pp. 44--60.

\bibitem{aws-fargate}
\BIBentryALTinterwordspacing
Amazon, ``Aws fargate,'' 2023. [Online]. Available:
  \url{https://docs.aws.amazon.com/eks/latest/userguide/fargate.html}
\BIBentrySTDinterwordspacing

\bibitem{google-kubernetes-engine}
\BIBentryALTinterwordspacing
Google, ``Google kubernetes engine (gke),'' 2023. [Online]. Available:
  \url{https://cloud.google.com/kubernetes-engine}
\BIBentrySTDinterwordspacing

\bibitem{azure-kubernetes-service}
\BIBentryALTinterwordspacing
Microsoft, ``Azure kubernetes service (ake),'' 2023. [Online]. Available:
  \url{https://azure.microsoft.com/en-us/products/kubernetes-service}
\BIBentrySTDinterwordspacing

\bibitem{aws-fargate-pricing}
\BIBentryALTinterwordspacing
Amazon, ``Aws fargate pricing,'' 2023. [Online]. Available:
  \url{https://aws.amazon.com/fargate/pricing/?nc1=h_ls}
\BIBentrySTDinterwordspacing

\bibitem{dubbo-12391}
\BIBentryALTinterwordspacing
Agorguy, ``dubbo pr-12391,'' 2023. [Online]. Available:
  \url{https://github.com/apache/dubbo/pull/12391}
\BIBentrySTDinterwordspacing

\bibitem{delight-nashorn-sandbox-137}
\BIBentryALTinterwordspacing
------, ``delight-nashorn-sandbox issue-137,'' 2023. [Online]. Available:
  \url{https://github.com/javadelight/delight-nashorn-sandbox/issues/137}
\BIBentrySTDinterwordspacing

\bibitem{http-core-747-issue}
\BIBentryALTinterwordspacing
------, ``http-core issue-747,'' 2023. [Online]. Available:
  \url{https://issues.apache.org/jira/browse/HTTPCORE-747}
\BIBentrySTDinterwordspacing

\bibitem{db-scheduler-388-issue}
\BIBentryALTinterwordspacing
------, ``db-scheduler issue-388,'' 2023. [Online]. Available:
  \url{https://github.com/kagkarlsson/db-scheduler/issues/388}
\BIBentrySTDinterwordspacing

\bibitem{timely-242}
\BIBentryALTinterwordspacing
------, ``timely issue-242,'' 2023. [Online]. Available:
  \url{https://github.com/NationalSecurityAgency/timely/issues/242}
\BIBentrySTDinterwordspacing

\bibitem{shardingsphere-elasticjob-2219}
\BIBentryALTinterwordspacing
------, ``shardingsphere-elasticjob issue-2219,'' 2023. [Online]. Available:
  \url{https://github.com/apache/shardingsphere-elasticjob/issues/2219}
\BIBentrySTDinterwordspacing

\bibitem{http-core-410}
\BIBentryALTinterwordspacing
------, ``http-core pr-410,'' 2023. [Online]. Available:
  \url{https://github.com/apache/httpcomponents-core/pull/410}
\BIBentrySTDinterwordspacing

\bibitem{db-scheduler-403}
\BIBentryALTinterwordspacing
------, ``db-scheduler pr-403,'' 2023. [Online]. Available:
  \url{https://github.com/kagkarlsson/db-scheduler/pull/403}
\BIBentrySTDinterwordspacing

\bibitem{java-websockets-test677}
\BIBentryALTinterwordspacing
N.~Rajlich, ``java-websockets issue677test,'' 2023. [Online]. Available:
  \url{https://github.com/TooTallNate/Java-WebSocket/blob/fa3909c391195178ccf5a92d4ac342a30ae247c8/src/test/java/org/java_websocket/issues/Issue677Test.java}
\BIBentrySTDinterwordspacing

\bibitem{GitHubTimely}
\BIBentryALTinterwordspacing
``{Timely},'' {A}ccessed 2023. [Online]. Available:
  \url{https://github.com/NationalSecurityAgency/timely}
\BIBentrySTDinterwordspacing

\bibitem{Lam2020WASM}
W.~Lam, S.~Winter, A.~Astorga, V.~Stodden, and D.~Marinov, ``Understanding
  reproducibility and characteristics of flaky tests through test reruns in
  {J}ava projects,'' in \emph{ISSRE}, 2020.

\bibitem{Silva2020TA}
D.~Silva, L.~Teixeira, and M.~d’Amorim, ``Shake it! {D}etecting flaky tests
  caused by concurrency with {S}haker,'' in \emph{ICSME}, 2020.

\bibitem{Gruber2022F}
M.~Gruber and G.~Fraser, ``A survey on how test flakiness affects developers
  and what support they need to address it,'' in \emph{ICST}, 2022.

\bibitem{awsCodeBuild}
{Amazon Web Services}, ``Aws codebuild pricing,''
  \url{https://aws.amazon.com/codebuild/pricing/}, 2023.

\bibitem{azureDevOps}
{Microsoft}, ``Pricing for azure devops,''
  \url{https://azure.microsoft.com/en-us/pricing/details/devops/azure-devops-services/},
  2023.

\bibitem{googleCloudBuild}
{Google Cloud Build}, ``Cloud build pricing,''
  \url{https://cloud.google.com/build/pricing}, 2023.

\bibitem{gitHubActionsRunners}
{GitHub}, ``About github-hosted runners,''
  \url{https://docs.github.com/en/actions/using-github-hosted-runners/about-github-hosted-runners},
  2023.

\bibitem{gitLabRunners}
{GitLab}, ``Saas runners on linux,''
  \url{https://docs.gitlab.com/ee/ci/runners/saas/linux_saas_runner.html},
  2023.

\bibitem{bitBucketRunners}
{Atlassian}, ``Bitbucket support - runners,''
  \url{https://support.atlassian.com/bitbucket-cloud/docs/runners/}, 2023.

\bibitem{circleCIRunners}
{CircleCI}, ``Configuring circleci,''
  \url{https://circleci.com/docs/configuration-reference/}, 2023.

\bibitem{travisBuild}
{TravisCI}, ``Travisci - build environment overview,''
  \url{https://docs.travis-ci.com/user/reference/overview/}, 2023.

\bibitem{teamCity}
{JetBrains}, ``Teamcity cloud,''
  \url{https://www.jetbrains.com/teamcity/cloud/}, 2023.

\bibitem{githubActionsAutoScale}
{GitHub}, ``Autoscaling with self-hosted runners,''
  \url{https://docs.github.com/en/actions/hosting-your-own-runners/autoscaling-with-self-hosted-runners},
  2023.

\bibitem{BenjaminiHochberg}
\BIBentryALTinterwordspacing
Y.~Benjamini and Y.~Hochberg, ``{Controlling the False Discovery Rate: A
  Practical and Powerful Approach to Multiple Testing},'' \emph{Journal of the
  Royal Statistical Society. Series B (Methodological)}, vol.~57, no.~1, pp.
  289--300, 1995. [Online]. Available:
  \url{http://www.jstor.org/stable/2346101}
\BIBentrySTDinterwordspacing

\end{thebibliography}

{\appendices
\onecolumn
\begin{landscape}
\section{RQ3 Appendix: Price and reliability per-project per-configuration}
\small
\setlength{\tabcolsep}{1pt}
\begin{longtable}{llllllllllllllllllllllllll}
\caption{For each of the cloud configurations, we report the number of builds (test suite executions) that contained at least one flaky test failure (F), and the cost per thousand builds in USD (P). The configuration with the lowest cost is bolded.}\\
& & \multicolumn{2}{c}{0.1c/1GiB} & \multicolumn{2}{c}{0.1c/2GiB} & \multicolumn{2}{c}{0.25c/2GiB} & \multicolumn{2}{c}{0.5c/2GiB} & \multicolumn{2}{c}{0.5c/4GiB} & \multicolumn{2}{c}{1c/4GiB} & \multicolumn{2}{c}{1c/8GiB} & \multicolumn{2}{c}{2c/4GiB} & \multicolumn{2}{c}{2c/8GiB} & \multicolumn{2}{c}{2c/16GiB} & \multicolumn{2}{c}{4c/8GiB} & \multicolumn{2}{c}{4c/16GiB} \\
\cmidrule(l{3pt}r{3pt}){3-4} \cmidrule(l{3pt}r{3pt}){5-6} \cmidrule(l{3pt}r{3pt}){7-8} \cmidrule(l{3pt}r{3pt}){9-10} \cmidrule(l{3pt}r{3pt}){11-12} \cmidrule(l{3pt}r{3pt}){13-14} \cmidrule(l{3pt}r{3pt}){15-16} \cmidrule(l{3pt}r{3pt}){17-18} \cmidrule(l{3pt}r{3pt}){19-20} \cmidrule(l{3pt}r{3pt}){21-22} \cmidrule(l{3pt}r{3pt}){23-24} \cmidrule(l{3pt}r{3pt}){25-26}
 \multicolumn{1}{c}{Language} & \multicolumn{1}{c}{Project}  & F & P & F & P & F & P & F & P & F & P & F & P & F & P & F & P & F & P & F & P & F & P & F & P\\
\midrule
\endhead
\cellcolor{gray!6}{java} & \cellcolor{gray!6}{assertj-core} & \cellcolor{gray!6}{229} & \cellcolor{gray!6}{0.16} & \cellcolor{gray!6}{202} & \cellcolor{gray!6}{0.24} & \cellcolor{gray!6}{57} & \cellcolor{gray!6}{0.13} & \cellcolor{gray!6}{27} & \cellcolor{gray!6}{0.10} & \cellcolor{gray!6}{26} & \cellcolor{gray!6}{0.12} & \cellcolor{gray!6}{29} & \cellcolor{gray!6}{0.10} & \cellcolor{gray!6}{32} & \cellcolor{gray!6}{0.13} & \cellcolor{gray!6}{29} & \cellcolor{gray!6}{\textbf{0.09}} & \cellcolor{gray!6}{23} & \cellcolor{gray!6}{0.11} & \cellcolor{gray!6}{41} & \cellcolor{gray!6}{0.18} & \cellcolor{gray!6}{34} & \cellcolor{gray!6}{0.15} & \cellcolor{gray!6}{32} & \cellcolor{gray!6}{0.19}\\
\cellcolor{gray!6}{java} & \cellcolor{gray!6}{assertj-core} & \cellcolor{gray!6}{229} & \cellcolor{gray!6}{0.16} & \cellcolor{gray!6}{202} & \cellcolor{gray!6}{0.24} & \cellcolor{gray!6}{57} & \cellcolor{gray!6}{0.13} & \cellcolor{gray!6}{27} & \cellcolor{gray!6}{0.10} & \cellcolor{gray!6}{26} & \cellcolor{gray!6}{0.12} & \cellcolor{gray!6}{29} & \cellcolor{gray!6}{0.10} & \cellcolor{gray!6}{32} & \cellcolor{gray!6}{0.13} & \cellcolor{gray!6}{29} & \cellcolor{gray!6}{\textbf{0.09}} & \cellcolor{gray!6}{23} & \cellcolor{gray!6}{0.11} & \cellcolor{gray!6}{41} & \cellcolor{gray!6}{0.18} & \cellcolor{gray!6}{34} & \cellcolor{gray!6}{0.15} & \cellcolor{gray!6}{32} & \cellcolor{gray!6}{0.19}\\
java & carbon-apimgt & 3 & \textbf{0.11} & 1 & 0.16 & 0 & \textbf{0.11} & 0 & \textbf{0.11} & 0 & 0.15 & 0 & 0.17 & 1 & 0.22 & 0 & 0.24 & 0 & 0.28 & 0 & 0.38 & 0 & 0.48 & 0 & 0.56\\
\cellcolor{gray!6}{java} & \cellcolor{gray!6}{commons-exec} & \cellcolor{gray!6}{11} & \cellcolor{gray!6}{\textbf{0.09}} & \cellcolor{gray!6}{6} & \cellcolor{gray!6}{0.13} & \cellcolor{gray!6}{6} & \cellcolor{gray!6}{0.12} & \cellcolor{gray!6}{8} & \cellcolor{gray!6}{0.16} & \cellcolor{gray!6}{6} & \cellcolor{gray!6}{0.21} & \cellcolor{gray!6}{2} & \cellcolor{gray!6}{0.29} & \cellcolor{gray!6}{4} & \cellcolor{gray!6}{0.37} & \cellcolor{gray!6}{0} & \cellcolor{gray!6}{0.45} & \cellcolor{gray!6}{1} & \cellcolor{gray!6}{0.55} & \cellcolor{gray!6}{0} & \cellcolor{gray!6}{0.72} & \cellcolor{gray!6}{0} & \cellcolor{gray!6}{0.97} & \cellcolor{gray!6}{0} & \cellcolor{gray!6}{1.11}\\
java & db-scheduler & 79 & 0.07 & 86 & 0.10 & 45 & \textbf{0.06} & 10 & \textbf{0.06} & 10 & 0.08 & 1 & 0.09 & 0 & 0.11 & 1 & 0.12 & 0 & 0.15 & 0 & 0.19 & 0 & 0.23 & 0 & 0.27\\
\cellcolor{gray!6}{java} & \cellcolor{gray!6}{delight-nashorn-sandbox} & \cellcolor{gray!6}{111} & \cellcolor{gray!6}{0.35} & \cellcolor{gray!6}{115} & \cellcolor{gray!6}{0.54} & \cellcolor{gray!6}{144} & \cellcolor{gray!6}{0.25} & \cellcolor{gray!6}{103} & \cellcolor{gray!6}{0.19} & \cellcolor{gray!6}{106} & \cellcolor{gray!6}{0.23} & \cellcolor{gray!6}{23} & \cellcolor{gray!6}{0.17} & \cellcolor{gray!6}{18} & \cellcolor{gray!6}{0.23} & \cellcolor{gray!6}{2} & \cellcolor{gray!6}{\textbf{0.16}} & \cellcolor{gray!6}{1} & \cellcolor{gray!6}{0.19} & \cellcolor{gray!6}{1} & \cellcolor{gray!6}{0.25} & \cellcolor{gray!6}{0} & \cellcolor{gray!6}{0.30} & \cellcolor{gray!6}{2} & \cellcolor{gray!6}{0.35}\\
java & elastic-job-lite & 0 & 0.28 & 0 & 0.42 & 0 & 0.25 & 0 & \textbf{0.24} & - & - & 0 & 0.29 & 0 & 0.38 & 0 & 0.49 & 0 & 0.49 & 0 & 0.65 & 0 & 0.97 & 0 & 0.89\\
\cellcolor{gray!6}{java} & \cellcolor{gray!6}{esper} & \cellcolor{gray!6}{0} & \cellcolor{gray!6}{0.07} & \cellcolor{gray!6}{0} & \cellcolor{gray!6}{0.11} & \cellcolor{gray!6}{0} & \cellcolor{gray!6}{0.07} & \cellcolor{gray!6}{2} & \cellcolor{gray!6}{\textbf{0.06}} & \cellcolor{gray!6}{1} & \cellcolor{gray!6}{0.07} & \cellcolor{gray!6}{4} & \cellcolor{gray!6}{0.07} & \cellcolor{gray!6}{3} & \cellcolor{gray!6}{0.10} & \cellcolor{gray!6}{22} & \cellcolor{gray!6}{0.09} & \cellcolor{gray!6}{17} & \cellcolor{gray!6}{0.10} & \cellcolor{gray!6}{16} & \cellcolor{gray!6}{0.13} & \cellcolor{gray!6}{19} & \cellcolor{gray!6}{0.16} & \cellcolor{gray!6}{26} & \cellcolor{gray!6}{0.19}\\
java & excelastic & 27 & 0.06 & 37 & 0.09 & 23 & 0.06 & 48 & \textbf{0.05} & 66 & 0.06 & 170 & 0.06 & 174 & 0.08 & 182 & 0.08 & 194 & 0.09 & 190 & 0.10 & 208 & 0.13 & 203 & 0.17\\
\cellcolor{gray!6}{java} & \cellcolor{gray!6}{fastjson} & \cellcolor{gray!6}{30} & \cellcolor{gray!6}{0.38} & \cellcolor{gray!6}{27} & \cellcolor{gray!6}{0.59} & \cellcolor{gray!6}{11} & \cellcolor{gray!6}{0.33} & \cellcolor{gray!6}{5} & \cellcolor{gray!6}{\textbf{0.27}} & \cellcolor{gray!6}{4} & \cellcolor{gray!6}{0.35} & \cellcolor{gray!6}{3} & \cellcolor{gray!6}{0.28} & \cellcolor{gray!6}{3} & \cellcolor{gray!6}{0.36} & \cellcolor{gray!6}{0} & \cellcolor{gray!6}{0.31} & \cellcolor{gray!6}{0} & \cellcolor{gray!6}{0.36} & \cellcolor{gray!6}{0} & \cellcolor{gray!6}{0.47} & \cellcolor{gray!6}{0} & \cellcolor{gray!6}{0.69} & \cellcolor{gray!6}{0} & \cellcolor{gray!6}{0.72}\\
java & fluent-logger-java & 114 & \textbf{0.09} & 61 & 0.14 & 1 & 0.10 & 0 & 0.12 & 0 & 0.15 & 0 & 0.19 & 0 & 0.25 & 0 & 0.30 & 0 & 0.36 & 0 & 0.46 & 0 & 0.59 & 0 & 0.70\\
\cellcolor{gray!6}{java} & \cellcolor{gray!6}{handlebars.java} & \cellcolor{gray!6}{10} & \cellcolor{gray!6}{0.12} & \cellcolor{gray!6}{6} & \cellcolor{gray!6}{0.18} & \cellcolor{gray!6}{6} & \cellcolor{gray!6}{0.10} & \cellcolor{gray!6}{1} & \cellcolor{gray!6}{0.08} & \cellcolor{gray!6}{3} & \cellcolor{gray!6}{0.10} & \cellcolor{gray!6}{2} & \cellcolor{gray!6}{0.08} & \cellcolor{gray!6}{8} & \cellcolor{gray!6}{0.10} & \cellcolor{gray!6}{5} & \cellcolor{gray!6}{\textbf{0.07}} & \cellcolor{gray!6}{3} & \cellcolor{gray!6}{0.09} & \cellcolor{gray!6}{5} & \cellcolor{gray!6}{0.11} & \cellcolor{gray!6}{7} & \cellcolor{gray!6}{0.13} & \cellcolor{gray!6}{5} & \cellcolor{gray!6}{0.17}\\
java & hector & 147 & 0.40 & 142 & 0.59 & 109 & 0.32 & 110 & \textbf{0.24} & 107 & 0.32 & 89 & 0.25 & 110 & 0.32 & 39 & 0.27 & 10 & 0.32 & 60 & 0.43 & 10 & 0.54 & 58 & 0.68\\
\cellcolor{gray!6}{java} & \cellcolor{gray!6}{http-request} & \cellcolor{gray!6}{0} & \cellcolor{gray!6}{0.05} & \cellcolor{gray!6}{0} & \cellcolor{gray!6}{0.08} & \cellcolor{gray!6}{0} & \cellcolor{gray!6}{0.04} & \cellcolor{gray!6}{0} & \cellcolor{gray!6}{\textbf{0.03}} & \cellcolor{gray!6}{0} & \cellcolor{gray!6}{0.04} & \cellcolor{gray!6}{0} & \cellcolor{gray!6}{\textbf{0.03}} & \cellcolor{gray!6}{0} & \cellcolor{gray!6}{0.04} & \cellcolor{gray!6}{0} & \cellcolor{gray!6}{\textbf{0.03}} & \cellcolor{gray!6}{0} & \cellcolor{gray!6}{0.04} & \cellcolor{gray!6}{0} & \cellcolor{gray!6}{0.05} & \cellcolor{gray!6}{0} & \cellcolor{gray!6}{0.07} & \cellcolor{gray!6}{0} & \cellcolor{gray!6}{0.08}\\
java & httpcore & 46 & 0.11 & 42 & 0.18 & 24 & 0.10 & 24 & \textbf{0.09} & 23 & 0.12 & 25 & 0.13 & 17 & 0.22 & 7 & 0.15 & 15 & 0.17 & 19 & 0.23 & 22 & 0.48 & 23 & 0.52\\
\cellcolor{gray!6}{java} & \cellcolor{gray!6}{hutool} & \cellcolor{gray!6}{4} & \cellcolor{gray!6}{0.05} & \cellcolor{gray!6}{2} & \cellcolor{gray!6}{0.06} & \cellcolor{gray!6}{2} & \cellcolor{gray!6}{0.04} & \cellcolor{gray!6}{4} & \cellcolor{gray!6}{0.03} & \cellcolor{gray!6}{1} & \cellcolor{gray!6}{0.04} & \cellcolor{gray!6}{1} & \cellcolor{gray!6}{0.03} & \cellcolor{gray!6}{1} & \cellcolor{gray!6}{0.04} & \cellcolor{gray!6}{0} & \cellcolor{gray!6}{\textbf{0.02}} & \cellcolor{gray!6}{0} & \cellcolor{gray!6}{0.04} & \cellcolor{gray!6}{1} & \cellcolor{gray!6}{0.05} & \cellcolor{gray!6}{0} & \cellcolor{gray!6}{0.05} & \cellcolor{gray!6}{1} & \cellcolor{gray!6}{0.06}\\
java & incubator-dubbo & - & - & - & - & 0 & 2.07 & 11 & \textbf{1.93} & 11 & 2.50 & 13 & 2.62 & 10 & 3.55 & 2 & 3.55 & 8 & 4.19 & 4 & 5.46 & 2 & 7.19 & 5 & 8.14\\
\cellcolor{gray!6}{java} & \cellcolor{gray!6}{java-websocket} & \cellcolor{gray!6}{12} & \cellcolor{gray!6}{0.10} & \cellcolor{gray!6}{10} & \cellcolor{gray!6}{0.14} & \cellcolor{gray!6}{73} & \cellcolor{gray!6}{0.09} & \cellcolor{gray!6}{91} & \cellcolor{gray!6}{\textbf{0.07}} & \cellcolor{gray!6}{104} & \cellcolor{gray!6}{0.09} & \cellcolor{gray!6}{99} & \cellcolor{gray!6}{0.08} & \cellcolor{gray!6}{95} & \cellcolor{gray!6}{0.10} & \cellcolor{gray!6}{97} & \cellcolor{gray!6}{0.08} & \cellcolor{gray!6}{111} & \cellcolor{gray!6}{0.12} & \cellcolor{gray!6}{99} & \cellcolor{gray!6}{0.14} & \cellcolor{gray!6}{87} & \cellcolor{gray!6}{0.15} & \cellcolor{gray!6}{61} & \cellcolor{gray!6}{0.23}\\
java & logback & 111 & \textbf{0.21} & 122 & 0.36 & 48 & 0.45 & 42 & 0.59 & 29 & 0.91 & 40 & 1.04 & 35 & 1.44 & 7 & 1.67 & 6 & 2.61 & 13 & 3.09 & 9 & 3.83 & 12 & 5.19\\
\cellcolor{gray!6}{java} & \cellcolor{gray!6}{luwak} & \cellcolor{gray!6}{8} & \cellcolor{gray!6}{0.12} & \cellcolor{gray!6}{13} & \cellcolor{gray!6}{0.18} & \cellcolor{gray!6}{12} & \cellcolor{gray!6}{0.10} & \cellcolor{gray!6}{3} & \cellcolor{gray!6}{\textbf{0.08}} & \cellcolor{gray!6}{1} & \cellcolor{gray!6}{0.10} & \cellcolor{gray!6}{0} & \cellcolor{gray!6}{0.09} & \cellcolor{gray!6}{12} & \cellcolor{gray!6}{0.11} & \cellcolor{gray!6}{0} & \cellcolor{gray!6}{0.11} & \cellcolor{gray!6}{0} & \cellcolor{gray!6}{0.12} & \cellcolor{gray!6}{1} & \cellcolor{gray!6}{0.15} & \cellcolor{gray!6}{2} & \cellcolor{gray!6}{0.20} & \cellcolor{gray!6}{0} & \cellcolor{gray!6}{0.21}\\
java & ninja & 4 & 0.23 & 5 & 0.35 & 8 & 0.19 & 6 & \textbf{0.15} & 7 & 0.20 & 7 & 0.17 & 8 & 0.21 & 4 & 0.18 & 11 & 0.19 & 11 & 0.25 & 11 & 0.30 & 7 & 0.33\\
\cellcolor{gray!6}{java} & \cellcolor{gray!6}{noxy} & \cellcolor{gray!6}{5} & \cellcolor{gray!6}{0.26} & \cellcolor{gray!6}{2} & \cellcolor{gray!6}{0.39} & \cellcolor{gray!6}{0} & \cellcolor{gray!6}{0.23} & \cellcolor{gray!6}{0} & \cellcolor{gray!6}{\textbf{0.16}} & \cellcolor{gray!6}{0} & \cellcolor{gray!6}{0.22} & \cellcolor{gray!6}{0} & \cellcolor{gray!6}{0.17} & \cellcolor{gray!6}{0} & \cellcolor{gray!6}{0.23} & \cellcolor{gray!6}{0} & \cellcolor{gray!6}{\textbf{0.16}} & \cellcolor{gray!6}{0} & \cellcolor{gray!6}{0.19} & \cellcolor{gray!6}{0} & \cellcolor{gray!6}{0.24} & \cellcolor{gray!6}{0} & \cellcolor{gray!6}{0.30} & \cellcolor{gray!6}{0} & \cellcolor{gray!6}{0.37}\\
java & orbit & 109 & \textbf{0.22} & 110 & 0.35 & 72 & 0.26 & 37 & 0.27 & 48 & 0.35 & 38 & 0.43 & 35 & 0.56 & 24 & 0.65 & 25 & 0.77 & 15 & 0.99 & 22 & 1.28 & 24 & 1.50\\
\cellcolor{gray!6}{java} & \cellcolor{gray!6}{oryx} & \cellcolor{gray!6}{0} & \cellcolor{gray!6}{\textbf{0.09}} & \cellcolor{gray!6}{0} & \cellcolor{gray!6}{0.15} & \cellcolor{gray!6}{0} & \cellcolor{gray!6}{0.10} & \cellcolor{gray!6}{0} & \cellcolor{gray!6}{0.11} & \cellcolor{gray!6}{0} & \cellcolor{gray!6}{0.14} & \cellcolor{gray!6}{0} & \cellcolor{gray!6}{0.17} & \cellcolor{gray!6}{0} & \cellcolor{gray!6}{0.22} & \cellcolor{gray!6}{1} & \cellcolor{gray!6}{0.26} & \cellcolor{gray!6}{3} & \cellcolor{gray!6}{0.31} & \cellcolor{gray!6}{0} & \cellcolor{gray!6}{0.41} & \cellcolor{gray!6}{2} & \cellcolor{gray!6}{0.48} & \cellcolor{gray!6}{1} & \cellcolor{gray!6}{0.58}\\
java & riptide & 0 & 0.12 & 0 & 0.19 & 0 & 0.12 & 0 & \textbf{0.11} & 0 & 0.15 & 0 & 0.16 & 0 & 0.20 & 0 & 0.21 & 8 & 0.29 & 0 & 0.33 & 0 & 0.26 & 0 & 0.31\\
\cellcolor{gray!6}{java} & \cellcolor{gray!6}{rxjava2-extras} & \cellcolor{gray!6}{3} & \cellcolor{gray!6}{0.22} & \cellcolor{gray!6}{1} & \cellcolor{gray!6}{0.35} & \cellcolor{gray!6}{1} & \cellcolor{gray!6}{0.22} & \cellcolor{gray!6}{1} & \cellcolor{gray!6}{\textbf{0.21}} & \cellcolor{gray!6}{1} & \cellcolor{gray!6}{0.27} & \cellcolor{gray!6}{6} & \cellcolor{gray!6}{0.30} & \cellcolor{gray!6}{1} & \cellcolor{gray!6}{0.39} & \cellcolor{gray!6}{2} & \cellcolor{gray!6}{0.41} & \cellcolor{gray!6}{3} & \cellcolor{gray!6}{0.50} & \cellcolor{gray!6}{2} & \cellcolor{gray!6}{0.65} & \cellcolor{gray!6}{1} & \cellcolor{gray!6}{0.87} & \cellcolor{gray!6}{5} & \cellcolor{gray!6}{0.95}\\
java & spring-boot & - & - & 1 & 1.69 & 0 & 1.02 & 1 & \textbf{0.73} & 2 & 0.97 & 2 & 0.80 & 1 & 0.98 & 1 & 0.77 & 0 & 0.90 & 0 & 1.25 & 1 & 1.46 & 0 & 2.00\\
\cellcolor{gray!6}{java} & \cellcolor{gray!6}{timely} & \cellcolor{gray!6}{0} & \cellcolor{gray!6}{0.22} & \cellcolor{gray!6}{0} & \cellcolor{gray!6}{0.33} & \cellcolor{gray!6}{0} & \cellcolor{gray!6}{0.18} & \cellcolor{gray!6}{0} & \cellcolor{gray!6}{0.14} & \cellcolor{gray!6}{0} & \cellcolor{gray!6}{0.17} & \cellcolor{gray!6}{0} & \cellcolor{gray!6}{0.13} & \cellcolor{gray!6}{0} & \cellcolor{gray!6}{0.17} & \cellcolor{gray!6}{0} & \cellcolor{gray!6}{\textbf{0.12}} & \cellcolor{gray!6}{0} & \cellcolor{gray!6}{0.14} & \cellcolor{gray!6}{0} & \cellcolor{gray!6}{0.18} & \cellcolor{gray!6}{0} & \cellcolor{gray!6}{0.20} & \cellcolor{gray!6}{0} & \cellcolor{gray!6}{0.23}\\
java & wro4j & - & - & - & - & - & - & - & - & 5 & 1.27 & 7 & 0.91 & 8 & 1.20 & 172 & \textbf{0.83} & 253 & 0.93 & 187 & 1.27 & 275 & 1.22 & 193 & 1.44\\
\cellcolor{gray!6}{java} & \cellcolor{gray!6}{yawp} & \cellcolor{gray!6}{46} & \cellcolor{gray!6}{0.08} & \cellcolor{gray!6}{53} & \cellcolor{gray!6}{0.13} & \cellcolor{gray!6}{11} & \cellcolor{gray!6}{0.07} & \cellcolor{gray!6}{9} & \cellcolor{gray!6}{\textbf{0.06}} & \cellcolor{gray!6}{6} & \cellcolor{gray!6}{0.07} & \cellcolor{gray!6}{8} & \cellcolor{gray!6}{\textbf{0.06}} & \cellcolor{gray!6}{14} & \cellcolor{gray!6}{0.08} & \cellcolor{gray!6}{5} & \cellcolor{gray!6}{0.07} & \cellcolor{gray!6}{17} & \cellcolor{gray!6}{0.09} & \cellcolor{gray!6}{14} & \cellcolor{gray!6}{0.11} & \cellcolor{gray!6}{23} & \cellcolor{gray!6}{0.13} & \cellcolor{gray!6}{23} & \cellcolor{gray!6}{0.16}\\
java & zxing & 29 & 0.67 & 22 & 0.90 & 20 & 0.50 & 23 & \textbf{0.38} & 24 & 0.49 & 23 & \textbf{0.38} & 17 & 0.49 & 25 & 0.56 & 18 & 0.66 & 27 & 0.86 & 18 & 1.12 & 22 & 1.32\\
\cellcolor{gray!6}{js} & \cellcolor{gray!6}{apollo-cache-persist} & \cellcolor{gray!6}{0} & \cellcolor{gray!6}{0.20} & \cellcolor{gray!6}{0} & \cellcolor{gray!6}{0.30} & \cellcolor{gray!6}{0} & \cellcolor{gray!6}{0.16} & \cellcolor{gray!6}{0} & \cellcolor{gray!6}{0.12} & \cellcolor{gray!6}{0} & \cellcolor{gray!6}{0.14} & \cellcolor{gray!6}{0} & \cellcolor{gray!6}{0.10} & \cellcolor{gray!6}{0} & \cellcolor{gray!6}{0.14} & \cellcolor{gray!6}{0} & \cellcolor{gray!6}{\textbf{0.09}} & \cellcolor{gray!6}{0} & \cellcolor{gray!6}{0.10} & \cellcolor{gray!6}{0} & \cellcolor{gray!6}{0.13} & \cellcolor{gray!6}{0} & \cellcolor{gray!6}{\textbf{0.09}} & \cellcolor{gray!6}{0} & \cellcolor{gray!6}{0.11}\\
js & apollo-client-devtools & 292 & 0.44 & 67 & 1.06 & 0 & 0.28 & 0 & \textbf{0.20} & 0 & 0.50 & 0 & 0.34 & 0 & 0.47 & 0 & 0.29 & 0 & 0.34 & 0 & 0.44 & 0 & 0.31 & 0 & 0.35\\
\cellcolor{gray!6}{js} & \cellcolor{gray!6}{AVA} & \cellcolor{gray!6}{0} & \cellcolor{gray!6}{1.81} & \cellcolor{gray!6}{0} & \cellcolor{gray!6}{1.34} & \cellcolor{gray!6}{0} & \cellcolor{gray!6}{0.67} & \cellcolor{gray!6}{0} & \cellcolor{gray!6}{0.50} & \cellcolor{gray!6}{0} & \cellcolor{gray!6}{0.61} & \cellcolor{gray!6}{0} & \cellcolor{gray!6}{0.41} & \cellcolor{gray!6}{0} & \cellcolor{gray!6}{0.58} & \cellcolor{gray!6}{0} & \cellcolor{gray!6}{\textbf{0.35}} & \cellcolor{gray!6}{0} & \cellcolor{gray!6}{0.41} & \cellcolor{gray!6}{0} & \cellcolor{gray!6}{0.52} & \cellcolor{gray!6}{0} & \cellcolor{gray!6}{0.37} & \cellcolor{gray!6}{0} & \cellcolor{gray!6}{0.42}\\
js & cavy & 0 & \textbf{0.01} & 0 & 0.02 & 0 & \textbf{0.01} & 0 & \textbf{0.01} & 0 & \textbf{0.01} & 0 & \textbf{0.01} & 0 & \textbf{0.01} & 0 & \textbf{0.01} & 0 & \textbf{0.01} & 0 & 0.02 & 0 & 0.02 & 0 & 0.03\\
\cellcolor{gray!6}{js} & \cellcolor{gray!6}{configure-aws-credentials} & \cellcolor{gray!6}{0} & \cellcolor{gray!6}{0.04} & \cellcolor{gray!6}{0} & \cellcolor{gray!6}{0.06} & \cellcolor{gray!6}{0} & \cellcolor{gray!6}{0.03} & \cellcolor{gray!6}{0} & \cellcolor{gray!6}{0.03} & \cellcolor{gray!6}{0} & \cellcolor{gray!6}{0.03} & \cellcolor{gray!6}{0} & \cellcolor{gray!6}{0.03} & \cellcolor{gray!6}{0} & \cellcolor{gray!6}{0.03} & \cellcolor{gray!6}{0} & \cellcolor{gray!6}{\textbf{0.02}} & \cellcolor{gray!6}{0} & \cellcolor{gray!6}{0.03} & \cellcolor{gray!6}{0} & \cellcolor{gray!6}{0.04} & \cellcolor{gray!6}{0} & \cellcolor{gray!6}{0.04} & \cellcolor{gray!6}{0} & \cellcolor{gray!6}{0.05}\\
js & derivablejs & 0 & 0.09 & 0 & 0.13 & 0 & 0.07 & 0 & 0.05 & 0 & 0.07 & 0 & \textbf{0.02} & 0 & 0.03 & 0 & 0.03 & 0 & 0.03 & 0 & 0.04 & 0 & 0.05 & 0 & 0.06\\
\cellcolor{gray!6}{js} & \cellcolor{gray!6}{facepaint} & \cellcolor{gray!6}{0} & \cellcolor{gray!6}{0.02} & \cellcolor{gray!6}{0} & \cellcolor{gray!6}{0.03} & \cellcolor{gray!6}{0} & \cellcolor{gray!6}{0.02} & \cellcolor{gray!6}{0} & \cellcolor{gray!6}{\textbf{0.01}} & \cellcolor{gray!6}{0} & \cellcolor{gray!6}{0.02} & \cellcolor{gray!6}{0} & \cellcolor{gray!6}{\textbf{0.01}} & \cellcolor{gray!6}{0} & \cellcolor{gray!6}{0.02} & \cellcolor{gray!6}{0} & \cellcolor{gray!6}{\textbf{0.01}} & \cellcolor{gray!6}{0} & \cellcolor{gray!6}{0.02} & \cellcolor{gray!6}{0} & \cellcolor{gray!6}{0.02} & \cellcolor{gray!6}{0} & \cellcolor{gray!6}{0.03} & \cellcolor{gray!6}{0} & \cellcolor{gray!6}{0.03}\\
js & graphdoc & 0 & 0.18 & 0 & 0.28 & 0 & 0.15 & 0 & 0.11 & 0 & 0.14 & 0 & 0.10 & 0 & 0.12 & 0 & \textbf{0.08} & 0 & 0.09 & 0 & 0.13 & 0 & 0.10 & 0 & 0.11\\
\cellcolor{gray!6}{js} & \cellcolor{gray!6}{IcedFrisby} & \cellcolor{gray!6}{300} & \cellcolor{gray!6}{0.03} & \cellcolor{gray!6}{1} & \cellcolor{gray!6}{0.02} & \cellcolor{gray!6}{0} & \cellcolor{gray!6}{\textbf{0.01}} & \cellcolor{gray!6}{0} & \cellcolor{gray!6}{\textbf{0.01}} & \cellcolor{gray!6}{0} & \cellcolor{gray!6}{0.02} & \cellcolor{gray!6}{26} & \cellcolor{gray!6}{0.03} & \cellcolor{gray!6}{300} & \cellcolor{gray!6}{0.15} & \cellcolor{gray!6}{300} & \cellcolor{gray!6}{0.20} & \cellcolor{gray!6}{300} & \cellcolor{gray!6}{0.23} & \cellcolor{gray!6}{300} & \cellcolor{gray!6}{0.30} & \cellcolor{gray!6}{300} & \cellcolor{gray!6}{0.39} & \cellcolor{gray!6}{0} & \cellcolor{gray!6}{0.05}\\
js & javascript-action & 0 & \textbf{0.01} & 0 & 0.02 & 0 & \textbf{0.01} & 0 & \textbf{0.01} & 0 & \textbf{0.01} & 0 & \textbf{0.01} & 0 & \textbf{0.01} & 0 & 0.02 & 0 & 0.02 & 0 & 0.02 & 0 & 0.03 & 0 & 0.04\\
\cellcolor{gray!6}{js} & \cellcolor{gray!6}{lock} & \cellcolor{gray!6}{0} & \cellcolor{gray!6}{0.68} & \cellcolor{gray!6}{0} & \cellcolor{gray!6}{0.95} & \cellcolor{gray!6}{0} & \cellcolor{gray!6}{0.54} & \cellcolor{gray!6}{0} & \cellcolor{gray!6}{0.40} & \cellcolor{gray!6}{0} & \cellcolor{gray!6}{0.50} & \cellcolor{gray!6}{0} & \cellcolor{gray!6}{\textbf{0.38}} & \cellcolor{gray!6}{0} & \cellcolor{gray!6}{0.50} & \cellcolor{gray!6}{0} & \cellcolor{gray!6}{0.51} & \cellcolor{gray!6}{0} & \cellcolor{gray!6}{0.63} & \cellcolor{gray!6}{0} & \cellcolor{gray!6}{0.80} & \cellcolor{gray!6}{0} & \cellcolor{gray!6}{1.02} & \cellcolor{gray!6}{0} & \cellcolor{gray!6}{1.11}\\
js & next-compose-plugins & 0 & 0.09 & 0 & 0.13 & 0 & 0.07 & 0 & 0.05 & 0 & 0.06 & 0 & 0.05 & 0 & 0.07 & 0 & \textbf{0.04} & 0 & 0.05 & 0 & 0.07 & 0 & 0.07 & 0 & 0.07\\
\cellcolor{gray!6}{js} & \cellcolor{gray!6}{ngrok} & \cellcolor{gray!6}{0} & \cellcolor{gray!6}{\textbf{0.02}} & \cellcolor{gray!6}{0} & \cellcolor{gray!6}{0.03} & \cellcolor{gray!6}{0} & \cellcolor{gray!6}{\textbf{0.02}} & \cellcolor{gray!6}{0} & \cellcolor{gray!6}{\textbf{0.02}} & \cellcolor{gray!6}{0} & \cellcolor{gray!6}{0.03} & \cellcolor{gray!6}{0} & \cellcolor{gray!6}{0.03} & \cellcolor{gray!6}{0} & \cellcolor{gray!6}{0.05} & \cellcolor{gray!6}{0} & \cellcolor{gray!6}{0.05} & \cellcolor{gray!6}{0} & \cellcolor{gray!6}{0.06} & \cellcolor{gray!6}{0} & \cellcolor{gray!6}{0.09} & \cellcolor{gray!6}{1} & \cellcolor{gray!6}{0.11} & \cellcolor{gray!6}{0} & \cellcolor{gray!6}{0.13}\\
js & preset-modules & 0 & 0.14 & 0 & 0.18 & 0 & 0.13 & 0 & 0.10 & 0 & 0.12 & 0 & 0.09 & 0 & 0.11 & 0 & \textbf{0.08} & 0 & 0.09 & 0 & 0.12 & 0 & 0.11 & 0 & 0.12\\
\cellcolor{gray!6}{js} & \cellcolor{gray!6}{react-datetime} & \cellcolor{gray!6}{0} & \cellcolor{gray!6}{0.15} & \cellcolor{gray!6}{0} & \cellcolor{gray!6}{0.22} & \cellcolor{gray!6}{0} & \cellcolor{gray!6}{0.13} & \cellcolor{gray!6}{0} & \cellcolor{gray!6}{\textbf{0.09}} & \cellcolor{gray!6}{0} & \cellcolor{gray!6}{0.10} & \cellcolor{gray!6}{0} & \cellcolor{gray!6}{0.10} & \cellcolor{gray!6}{0} & \cellcolor{gray!6}{0.11} & \cellcolor{gray!6}{0} & \cellcolor{gray!6}{\textbf{0.09}} & \cellcolor{gray!6}{0} & \cellcolor{gray!6}{0.11} & \cellcolor{gray!6}{0} & \cellcolor{gray!6}{0.14} & \cellcolor{gray!6}{0} & \cellcolor{gray!6}{0.13} & \cellcolor{gray!6}{0} & \cellcolor{gray!6}{0.15}\\
js & react-native & 26 & 1.01 & 21 & 1.79 & 2 & 0.92 & 1 & 0.68 & 0 & 0.79 & 0 & 0.57 & 0 & 0.77 & 0 & \textbf{0.46} & 0 & 0.56 & 0 & 0.74 & 0 & 1.19 & 0 & 0.64\\
\cellcolor{gray!6}{js} & \cellcolor{gray!6}{react-slick} & \cellcolor{gray!6}{0} & \cellcolor{gray!6}{0.34} & \cellcolor{gray!6}{0} & \cellcolor{gray!6}{0.51} & \cellcolor{gray!6}{0} & \cellcolor{gray!6}{0.27} & \cellcolor{gray!6}{0} & \cellcolor{gray!6}{0.19} & \cellcolor{gray!6}{0} & \cellcolor{gray!6}{0.25} & \cellcolor{gray!6}{0} & \cellcolor{gray!6}{0.17} & \cellcolor{gray!6}{0} & \cellcolor{gray!6}{0.23} & \cellcolor{gray!6}{0} & \cellcolor{gray!6}{\textbf{0.14}} & \cellcolor{gray!6}{0} & \cellcolor{gray!6}{0.18} & \cellcolor{gray!6}{0} & \cellcolor{gray!6}{0.23} & \cellcolor{gray!6}{0} & \cellcolor{gray!6}{0.16} & \cellcolor{gray!6}{0} & \cellcolor{gray!6}{0.19}\\
js & react-stdio & 0 & 0.02 & 0 & 0.02 & 0 & \textbf{0.01} & 0 & \textbf{0.01} & 0 & \textbf{0.01} & 0 & \textbf{0.01} & 0 & \textbf{0.01} & 0 & \textbf{0.01} & 0 & 0.02 & 0 & 0.02 & 0 & 0.03 & 0 & 0.03\\
\cellcolor{gray!6}{js} & \cellcolor{gray!6}{redux-ship} & \cellcolor{gray!6}{0} & \cellcolor{gray!6}{0.02} & \cellcolor{gray!6}{0} & \cellcolor{gray!6}{0.03} & \cellcolor{gray!6}{0} & \cellcolor{gray!6}{\textbf{0.01}} & \cellcolor{gray!6}{0} & \cellcolor{gray!6}{\textbf{0.01}} & \cellcolor{gray!6}{0} & \cellcolor{gray!6}{\textbf{0.01}} & \cellcolor{gray!6}{0} & \cellcolor{gray!6}{\textbf{0.01}} & \cellcolor{gray!6}{0} & \cellcolor{gray!6}{\textbf{0.01}} & \cellcolor{gray!6}{0} & \cellcolor{gray!6}{\textbf{0.01}} & \cellcolor{gray!6}{0} & \cellcolor{gray!6}{0.02} & \cellcolor{gray!6}{0} & \cellcolor{gray!6}{0.02} & \cellcolor{gray!6}{0} & \cellcolor{gray!6}{0.02} & \cellcolor{gray!6}{0} & \cellcolor{gray!6}{0.03}\\
js & resume-cli & 0 & 0.08 & 0 & 0.12 & 0 & \textbf{0.07} & 0 & \textbf{0.07} & 0 & 0.08 & 0 & 0.09 & 0 & 0.12 & 0 & 0.13 & 0 & 0.16 & 0 & 0.20 & 0 & 0.26 & 0 & 0.31\\
\cellcolor{gray!6}{js} & \cellcolor{gray!6}{semantic-pull-requests} & \cellcolor{gray!6}{0} & \cellcolor{gray!6}{0.05} & \cellcolor{gray!6}{0} & \cellcolor{gray!6}{0.08} & \cellcolor{gray!6}{0} & \cellcolor{gray!6}{0.04} & \cellcolor{gray!6}{0} & \cellcolor{gray!6}{\textbf{0.03}} & \cellcolor{gray!6}{0} & \cellcolor{gray!6}{0.04} & \cellcolor{gray!6}{0} & \cellcolor{gray!6}{\textbf{0.03}} & \cellcolor{gray!6}{0} & \cellcolor{gray!6}{0.04} & \cellcolor{gray!6}{0} & \cellcolor{gray!6}{0.04} & \cellcolor{gray!6}{0} & \cellcolor{gray!6}{0.04} & \cellcolor{gray!6}{0} & \cellcolor{gray!6}{0.05} & \cellcolor{gray!6}{0} & \cellcolor{gray!6}{0.07} & \cellcolor{gray!6}{0} & \cellcolor{gray!6}{0.08}\\
js & shields & 105 & 0.03 & 205 & 0.05 & 0 & 0.03 & 0 & 0.02 & 0 & 0.03 & 0 & \textbf{0.01} & 0 & \textbf{0.01} & 0 & 0.02 & 0 & 0.02 & 0 & 0.03 & 0 & 0.03 & 0 & 0.03\\
\cellcolor{gray!6}{js} & \cellcolor{gray!6}{tippyjs-react} & \cellcolor{gray!6}{0} & \cellcolor{gray!6}{0.23} & \cellcolor{gray!6}{0} & \cellcolor{gray!6}{0.34} & \cellcolor{gray!6}{0} & \cellcolor{gray!6}{0.18} & \cellcolor{gray!6}{0} & \cellcolor{gray!6}{0.13} & \cellcolor{gray!6}{0} & \cellcolor{gray!6}{0.16} & \cellcolor{gray!6}{0} & \cellcolor{gray!6}{0.12} & \cellcolor{gray!6}{0} & \cellcolor{gray!6}{0.17} & \cellcolor{gray!6}{0} & \cellcolor{gray!6}{\textbf{0.11}} & \cellcolor{gray!6}{0} & \cellcolor{gray!6}{0.13} & \cellcolor{gray!6}{0} & \cellcolor{gray!6}{0.17} & \cellcolor{gray!6}{0} & \cellcolor{gray!6}{0.15} & \cellcolor{gray!6}{0} & \cellcolor{gray!6}{0.16}\\
js & twilio-video-app-react & - & - & 16 & 2.58 & 3 & 1.41 & 0 & 1.08 & 0 & 0.99 & 0 & 0.74 & 0 & 1.03 & 0 & 0.89 & 0 & 0.75 & 0 & 0.94 & 0 & \textbf{0.68} & 0 & 0.77\\
\cellcolor{gray!6}{js} & \cellcolor{gray!6}{whenipress} & \cellcolor{gray!6}{0} & \cellcolor{gray!6}{0.04} & \cellcolor{gray!6}{0} & \cellcolor{gray!6}{0.06} & \cellcolor{gray!6}{0} & \cellcolor{gray!6}{\textbf{0.03}} & \cellcolor{gray!6}{0} & \cellcolor{gray!6}{\textbf{0.03}} & \cellcolor{gray!6}{0} & \cellcolor{gray!6}{0.04} & \cellcolor{gray!6}{0} & \cellcolor{gray!6}{\textbf{0.03}} & \cellcolor{gray!6}{0} & \cellcolor{gray!6}{0.04} & \cellcolor{gray!6}{0} & \cellcolor{gray!6}{\textbf{0.03}} & \cellcolor{gray!6}{0} & \cellcolor{gray!6}{0.04} & \cellcolor{gray!6}{0} & \cellcolor{gray!6}{0.05} & \cellcolor{gray!6}{0} & \cellcolor{gray!6}{0.05} & \cellcolor{gray!6}{0} & \cellcolor{gray!6}{0.06}\\
python & celery & 0 & \textbf{0.08} & 0 & 0.13 & 0 & 0.19 & 0 & 0.29 & 0 & 0.61 & 0 & 0.53 & 0 & 0.71 & 0 & 0.91 & 0 & 1.07 & 0 & 1.41 & 0 & 1.78 & 0 & 2.16\\
\cellcolor{gray!6}{python} & \cellcolor{gray!6}{conan} & \cellcolor{gray!6}{0} & \cellcolor{gray!6}{4.86} & \cellcolor{gray!6}{0} & \cellcolor{gray!6}{7.12} & \cellcolor{gray!6}{0} & \cellcolor{gray!6}{4.08} & \cellcolor{gray!6}{0} & \cellcolor{gray!6}{\textbf{3.13}} & \cellcolor{gray!6}{0} & \cellcolor{gray!6}{4.06} & \cellcolor{gray!6}{0} & \cellcolor{gray!6}{3.47} & \cellcolor{gray!6}{0} & \cellcolor{gray!6}{4.53} & \cellcolor{gray!6}{0} & \cellcolor{gray!6}{5.81} & \cellcolor{gray!6}{0} & \cellcolor{gray!6}{6.94} & \cellcolor{gray!6}{0} & \cellcolor{gray!6}{8.41} & \cellcolor{gray!6}{0} & \cellcolor{gray!6}{11.43} & \cellcolor{gray!6}{0} & \cellcolor{gray!6}{13.41}\\
python & django-rest-framework & 0 & 0.11 & 0 & 0.16 & 0 & 0.09 & 0 & \textbf{0.07} & 0 & 0.09 & 0 & \textbf{0.07} & 0 & 0.09 & 0 & 0.11 & 0 & 0.13 & 0 & 0.18 & 0 & 0.23 & 0 & 0.27\\
\cellcolor{gray!6}{python} & \cellcolor{gray!6}{electrum} & \cellcolor{gray!6}{0} & \cellcolor{gray!6}{0.51} & \cellcolor{gray!6}{0} & \cellcolor{gray!6}{0.78} & \cellcolor{gray!6}{0} & \cellcolor{gray!6}{0.47} & \cellcolor{gray!6}{0} & \cellcolor{gray!6}{\textbf{0.34}} & \cellcolor{gray!6}{0} & \cellcolor{gray!6}{0.46} & \cellcolor{gray!6}{0} & \cellcolor{gray!6}{0.40} & \cellcolor{gray!6}{0} & \cellcolor{gray!6}{0.53} & \cellcolor{gray!6}{0} & \cellcolor{gray!6}{0.68} & \cellcolor{gray!6}{0} & \cellcolor{gray!6}{0.80} & \cellcolor{gray!6}{0} & \cellcolor{gray!6}{1.09} & \cellcolor{gray!6}{0} & \cellcolor{gray!6}{1.35} & \cellcolor{gray!6}{0} & \cellcolor{gray!6}{1.62}\\
python & Flexget & 0 & 2.62 & 0 & 3.95 & 0 & 2.27 & 0 & \textbf{1.72} & 0 & 2.32 & 0 & 1.81 & 0 & 2.35 & 0 & 3.03 & 0 & 3.55 & 0 & 4.70 & 0 & 6.13 & 0 & 7.27\\
\cellcolor{gray!6}{python} & \cellcolor{gray!6}{fonttools} & \cellcolor{gray!6}{0} & \cellcolor{gray!6}{0.21} & \cellcolor{gray!6}{0} & \cellcolor{gray!6}{0.31} & \cellcolor{gray!6}{0} & \cellcolor{gray!6}{0.18} & \cellcolor{gray!6}{0} & \cellcolor{gray!6}{\textbf{0.13}} & \cellcolor{gray!6}{0} & \cellcolor{gray!6}{0.18} & \cellcolor{gray!6}{0} & \cellcolor{gray!6}{0.14} & \cellcolor{gray!6}{0} & \cellcolor{gray!6}{0.17} & \cellcolor{gray!6}{0} & \cellcolor{gray!6}{0.22} & \cellcolor{gray!6}{0} & \cellcolor{gray!6}{0.27} & \cellcolor{gray!6}{0} & \cellcolor{gray!6}{0.35} & \cellcolor{gray!6}{0} & \cellcolor{gray!6}{0.46} & \cellcolor{gray!6}{0} & \cellcolor{gray!6}{0.55}\\
python & graphene & 0 & \textbf{0.03} & 0 & 0.05 & 0 & 0.04 & 0 & \textbf{0.03} & 0 & 0.05 & 0 & 0.04 & 0 & 0.06 & 0 & 0.07 & 0 & 0.08 & 0 & 0.11 & 0 & 0.14 & 0 & 0.17\\
\cellcolor{gray!6}{python} & \cellcolor{gray!6}{hydra} & \cellcolor{gray!6}{0} & \cellcolor{gray!6}{0.90} & \cellcolor{gray!6}{0} & \cellcolor{gray!6}{1.41} & \cellcolor{gray!6}{0} & \cellcolor{gray!6}{0.79} & \cellcolor{gray!6}{0} & \cellcolor{gray!6}{\textbf{0.59}} & \cellcolor{gray!6}{0} & \cellcolor{gray!6}{0.77} & \cellcolor{gray!6}{0} & \cellcolor{gray!6}{\textbf{0.59}} & \cellcolor{gray!6}{0} & \cellcolor{gray!6}{0.77} & \cellcolor{gray!6}{0} & \cellcolor{gray!6}{0.98} & \cellcolor{gray!6}{0} & \cellcolor{gray!6}{1.17} & \cellcolor{gray!6}{0} & \cellcolor{gray!6}{1.52} & \cellcolor{gray!6}{0} & \cellcolor{gray!6}{1.97} & \cellcolor{gray!6}{0} & \cellcolor{gray!6}{2.22}\\
python & ipython & 0 & 0.23 & 0 & 0.34 & 0 & 0.22 & 0 & \textbf{0.20} & 0 & 0.26 & 0 & 0.26 & 0 & 0.34 & 0 & 0.44 & 0 & 0.52 & 0 & 0.68 & 0 & 0.88 & 0 & 1.05\\
\cellcolor{gray!6}{python} & \cellcolor{gray!6}{kombu} & \cellcolor{gray!6}{0} & \cellcolor{gray!6}{\textbf{0.08}} & \cellcolor{gray!6}{0} & \cellcolor{gray!6}{0.11} & \cellcolor{gray!6}{0} & \cellcolor{gray!6}{\textbf{0.08}} & \cellcolor{gray!6}{0} & \cellcolor{gray!6}{\textbf{0.08}} & \cellcolor{gray!6}{0} & \cellcolor{gray!6}{0.11} & \cellcolor{gray!6}{0} & \cellcolor{gray!6}{0.13} & \cellcolor{gray!6}{0} & \cellcolor{gray!6}{0.17} & \cellcolor{gray!6}{0} & \cellcolor{gray!6}{0.22} & \cellcolor{gray!6}{0} & \cellcolor{gray!6}{0.27} & \cellcolor{gray!6}{0} & \cellcolor{gray!6}{0.34} & \cellcolor{gray!6}{0} & \cellcolor{gray!6}{0.44} & \cellcolor{gray!6}{0} & \cellcolor{gray!6}{0.50}\\
python & loguru & 290 & \textbf{0.28} & 0 & 0.42 & 0 & 0.29 & 0 & \textbf{0.28} & 0 & 0.37 & 0 & 0.41 & 0 & 0.53 & 270 & 0.69 & 0 & 0.81 & 0 & 1.06 & 260 & 1.37 & 0 & 1.62\\
\cellcolor{gray!6}{python} & \cellcolor{gray!6}{mitmproxy} & \cellcolor{gray!6}{0} & \cellcolor{gray!6}{0.17} & \cellcolor{gray!6}{0} & \cellcolor{gray!6}{0.25} & \cellcolor{gray!6}{0} & \cellcolor{gray!6}{0.15} & \cellcolor{gray!6}{0} & \cellcolor{gray!6}{\textbf{0.12}} & \cellcolor{gray!6}{0} & \cellcolor{gray!6}{0.15} & \cellcolor{gray!6}{0} & \cellcolor{gray!6}{0.13} & \cellcolor{gray!6}{0} & \cellcolor{gray!6}{0.17} & \cellcolor{gray!6}{0} & \cellcolor{gray!6}{0.22} & \cellcolor{gray!6}{0} & \cellcolor{gray!6}{0.27} & \cellcolor{gray!6}{0} & \cellcolor{gray!6}{0.34} & \cellcolor{gray!6}{0} & \cellcolor{gray!6}{0.44} & \cellcolor{gray!6}{0} & \cellcolor{gray!6}{0.52}\\
python & Pillow & 0 & \textbf{0.04} & 0 & 0.07 & 0 & 0.10 & 0 & 0.15 & 0 & 0.35 & 0 & 0.28 & 0 & 0.35 & 0 & 0.45 & 0 & 0.53 & 0 & 0.67 & 0 & 0.90 & 0 & 1.06\\
\cellcolor{gray!6}{python} & \cellcolor{gray!6}{prefect} & \cellcolor{gray!6}{0} & \cellcolor{gray!6}{\textbf{4.40}} & \cellcolor{gray!6}{0} & \cellcolor{gray!6}{6.70} & \cellcolor{gray!6}{0} & \cellcolor{gray!6}{9.85} & \cellcolor{gray!6}{0} & \cellcolor{gray!6}{15.09} & \cellcolor{gray!6}{0} & \cellcolor{gray!6}{19.69} & \cellcolor{gray!6}{0} & \cellcolor{gray!6}{30.17} & \cellcolor{gray!6}{0} & \cellcolor{gray!6}{39.38} & \cellcolor{gray!6}{0} & \cellcolor{gray!6}{51.14} & \cellcolor{gray!6}{0} & \cellcolor{gray!6}{60.35} & \cellcolor{gray!6}{0} & \cellcolor{gray!6}{78.77} & \cellcolor{gray!6}{0} & \cellcolor{gray!6}{102.28} & \cellcolor{gray!6}{0} & \cellcolor{gray!6}{120.70}\\
python & PyGithub & 0 & 0.13 & 0 & 0.20 & 0 & 0.12 & 0 & \textbf{0.10} & 0 & 0.14 & 0 & 0.13 & 0 & 0.17 & 0 & 0.21 & 0 & 0.25 & 0 & 0.33 & 0 & 0.42 & 0 & 0.50\\
\cellcolor{gray!6}{python} & \cellcolor{gray!6}{pyramid} & \cellcolor{gray!6}{0} & \cellcolor{gray!6}{0.10} & \cellcolor{gray!6}{0} & \cellcolor{gray!6}{0.15} & \cellcolor{gray!6}{0} & \cellcolor{gray!6}{0.08} & \cellcolor{gray!6}{0} & \cellcolor{gray!6}{\textbf{0.06}} & \cellcolor{gray!6}{0} & \cellcolor{gray!6}{0.08} & \cellcolor{gray!6}{0} & \cellcolor{gray!6}{\textbf{0.06}} & \cellcolor{gray!6}{0} & \cellcolor{gray!6}{0.08} & \cellcolor{gray!6}{0} & \cellcolor{gray!6}{0.11} & \cellcolor{gray!6}{0} & \cellcolor{gray!6}{0.12} & \cellcolor{gray!6}{0} & \cellcolor{gray!6}{0.16} & \cellcolor{gray!6}{0} & \cellcolor{gray!6}{0.21} & \cellcolor{gray!6}{0} & \cellcolor{gray!6}{0.24}\\
python & requests & 23 & \textbf{0.09} & 29 & 0.13 & 19 & 0.17 & 19 & 0.25 & 21 & 0.33 & 18 & 0.49 & 21 & 0.65 & 19 & 0.84 & 20 & 0.99 & 23 & 1.27 & 24 & 1.63 & 19 & 2.00\\
\cellcolor{gray!6}{python} & \cellcolor{gray!6}{seaborn} & \cellcolor{gray!6}{0} & \cellcolor{gray!6}{\textbf{0.26}} & \cellcolor{gray!6}{0} & \cellcolor{gray!6}{4.42} & \cellcolor{gray!6}{0} & \cellcolor{gray!6}{2.85} & \cellcolor{gray!6}{0} & \cellcolor{gray!6}{2.64} & \cellcolor{gray!6}{0} & \cellcolor{gray!6}{3.46} & \cellcolor{gray!6}{0} & \cellcolor{gray!6}{3.47} & \cellcolor{gray!6}{0} & \cellcolor{gray!6}{4.48} & \cellcolor{gray!6}{0} & \cellcolor{gray!6}{4.67} & \cellcolor{gray!6}{0} & \cellcolor{gray!6}{5.55} & \cellcolor{gray!6}{0} & \cellcolor{gray!6}{7.36} & \cellcolor{gray!6}{0} & \cellcolor{gray!6}{8.58} & \cellcolor{gray!6}{0} & \cellcolor{gray!6}{10.23}\\
python & setuptools & 1 & 3.24 & 0 & 4.97 & 299 & 2.70 & 298 & \textbf{2.22} & 298 & 2.89 & 298 & 4.17 & 298 & 5.36 & 0 & 7.02 & 300 & 8.36 & 297 & 10.35 & 300 & 14.31 & 5 & 16.66\\
\cellcolor{gray!6}{python} & \cellcolor{gray!6}{sunpy} & \cellcolor{gray!6}{0} & \cellcolor{gray!6}{1.37} & \cellcolor{gray!6}{0} & \cellcolor{gray!6}{2.08} & \cellcolor{gray!6}{0} & \cellcolor{gray!6}{1.22} & \cellcolor{gray!6}{0} & \cellcolor{gray!6}{\textbf{0.96}} & \cellcolor{gray!6}{0} & \cellcolor{gray!6}{1.24} & \cellcolor{gray!6}{0} & \cellcolor{gray!6}{0.98} & \cellcolor{gray!6}{0} & \cellcolor{gray!6}{1.30} & \cellcolor{gray!6}{0} & \cellcolor{gray!6}{1.47} & \cellcolor{gray!6}{0} & \cellcolor{gray!6}{1.75} & \cellcolor{gray!6}{0} & \cellcolor{gray!6}{2.30} & \cellcolor{gray!6}{0} & \cellcolor{gray!6}{2.69} & \cellcolor{gray!6}{0} & \cellcolor{gray!6}{3.26}\\
python & xonsh & 0 & 0.63 & 0 & 0.98 & 0 & 0.52 & 0 & \textbf{0.40} & 0 & 0.51 & 0 & 0.41 & 0 & 0.52 & 0 & 0.66 & 0 & 0.79 & 0 & 1.04 & 0 & 1.35 & 0 & 1.57\\
\bottomrule
\end{longtable}
\end{landscape}
\twocolumn

}

\end{document}